\documentclass[trackchanges, twocolumn]{aastex701}

\newcommand{\cahk}{Ca \scriptsize{\uppercase\expandafter{\romannumeral2}} \normalsize H$\&$K }
\newcommand{\cai}{Ca \scriptsize{\uppercase\expandafter{\romannumeral1}} \normalsize }
\newcommand{\caii}{Ca \scriptsize{\uppercase\expandafter{\romannumeral2}} \normalsize }
\usepackage{amsmath}
\usepackage{chngcntr}
\usepackage[figuresright]{rotating}
\usepackage{ctable}
\usepackage{float}
\usepackage{adjustbox}
\usepackage{graphicx}
\usepackage{realboxes}
\usepackage{epsfig}
\usepackage{threeparttablex, tablefootnote}
\usepackage{subfigure}
\usepackage{graphicx}
\usepackage{CJK}
\usepackage{booktabs}
\usepackage{multirow}
\usepackage{amsmath}

\begin{document}

\title{Evolution of Stellar Activity and Habitable Zone II: \cahk Emissions of Late-type Dwarfs}

\author[orcid=0000-0003-3474-5118,gname=Henggeng, sname='Han']{Henggeng Han} 
\affiliation{National Astronomical Observatories, Chinese Academy of Sciences, Beijing 100101, People's Republic of China}
\email{hghan@nao.cas.cn}

\author[orcid=0000-0003-3116-5038,gname=Song, sname='Wang']{Song Wang} 
\altaffiliation{Corresponding author: songw@bao.ac.cn}
\affiliation{National Astronomical Observatories, Chinese Academy of Sciences, Beijing 100101, People's Republic of China}
\affiliation{Institute for Frontiers in Astronomy and Astrophysics, Beijing Normal University, Beijing, 102206, People's Republic of China}
\email{songw@bao.ac.cn}

\author[orcid=0009-0007-4501-4376,gname=Xue, sname='Li']{Xue Li} 
\affiliation{School of Astronomy and Space Science, University of Chinese Academy of Sciences, Beijing 100049, People's Republic of China}
\email{lixue@bao.ac.cn}

\author[orcid=0009-0006-7556-8401,gname=Chuanjie, sname='Zheng']{Chuanjie Zheng} 
\affiliation{School of Astronomy and Space Science, University of Chinese Academy of Sciences, Beijing 100049, People's Republic of China}
\email{zhengchuanjie@ucas.ac.cn}

\author[orcid=0009-0009-3372-5271,gname=Jiahui, sname='Wang']{Jiahui Wang} 
\affiliation{National Astronomical Observatories, Chinese Academy of Sciences, Beijing 100101, People's Republic of China}
\affiliation{School of Astronomy and Space Science, University of Chinese Academy of Sciences, Beijing 100049, People's Republic of China}
\email{wangjh@bao.ac.cn}

\author[orcid=0000-0002-2874-2706,gname=Jifeng, sname='Liu']{Jifeng Liu} 
\affiliation{National Astronomical Observatories, Chinese Academy of Sciences, Beijing 100101, People's Republic of China}
\affiliation{School of Astronomy and Space Science, University of Chinese Academy of Sciences, Beijing 100049, People's Republic of China}
\affiliation{Institute for Frontiers in Astronomy and Astrophysics, Beijing Normal University, Beijing, 102206, People's Republic of China}
\affiliation{New Cornerstone Science Laboratory, National Astronomical Observatories, Chinese Academy of Sciences, Beijing, 100012, People's Republic of China}
\email{jfliu@bao.ac.cn}


\begin{abstract}
Stellar chromospheric activity serves as a valuable proxy for estimating stellar ages, though its applicable range and accurate functional form are still debated.
In this study, utilizing the LAMOST spectra we compiled a catalog of open cluster members and field stars to investigate $R_{\rm{HK}}^{'}$--age relations across various spectral types. We find that a linear model, specifically a Skumanich-type relation, can best describe the overall decline of chromospheric activity with age, with the slope varying across different spectral types. However, we also identify variations in the decay rate along the main sequence, which call for more accurate follow-up investigation.
Finally, we find that lower-metallicity stars exhibit enhanced activity for F-, G-, and K-type stars, whereas no clear metallicity dependence is observed for M dwarfs.

\end{abstract}



\section{Introduction}
\label{sec:intro}
Late-type stars will spin down as they evolved due to the angular momentum loss caused by magnetized stellar winds \citep{1968MNRAS.138..359M, 1988ApJ...333..236K}. As a result, stellar rotation periods can serve as a diagnostic of stellar ages, an approach known as gyrochronology \citep{2003ApJ...586..464B}. Given the close connection between magnetic activity and rotation, magnetic activity is therefore also expected to evolve with age.

One of the pioneering studies on the evolution of stellar magnetic activity was conducted by \cite{1972ApJ...171..565S}, who proposed an inverse square-root law relating magnetic activity, rotation and Li abundance to stellar age. Later on, some discrepancies appeared. For example, \cite{1991ApJ...375..722S} applied the Skumanich-type relation to samples from \cite{1980PASP...92..385V} and found a different exponent, which they suggested will result in an overpopulation of young stars with ages below 1 Gyr. Meanwhile, they proposed that a curve could also fit the data quite well, corresponding to a constant star formation rates. 

As sample sizes continued to grown, increasing discrepancies with this simple picture continued to emerge. On the one hand, some studies still supported the Skumanich-type scenario. \cite{2018A&A...619A..73L} compiled an elite sample of solar twins and suggested the Skumanich-type relation may extend to $\sim$ 6 Gyr, a finding further supported by \cite{2025MNRAS.542.2431L} using open cluster members. Meanwhile, \cite{2024ApJS..271...19Y} calibrated activity--age relations using \emph{Kepler}-LAMOST overlap stars and proposed they remain valid for G- and K-type dwarfs up to $\sim$13 Gyr. 

Conversely, other studies argued that the evolutionary history was more complex. Some proposed an initial rapid decline in chromospheric activity followed by a phase in which activity level almost keeps unchanged \citep{2004A&A...426.1021P, 2009A&A...499L...9P}. Others have identified and modeled variable decay rates at different evolutionary stages \citep{2008ApJ...687.1264M, 2017MNRAS.471.1012B}. These works suggested that the activity--age relations may present variable decay rate. 

While activity--age relations have been extensively studied for open clusters, research on old field dwarfs has long been constrained by the lack of reliable age estimates for large samples. To systematically investigate the evolution of chromospheric activity across different stellar types and evolutionary phases, it is essential to combine both young open clusters and old field stars. Recently \cite{2025ApJS..280...13W} derived ages for $\sim$4 million LAMOST DR10 dwarfs. This large field star sample, together with existing open cluster observations, now enables the extension and calibration of the activity--age relation across a vast age range, which may enable a more universally applicable relation.

In our series paper, \cite{2025ApJS..281...13L} \label{cite:paper1} (hereafter Paper \ref{cite:paper1}), we explored the activity--age relations for F-, G-, K-, and M-type dwarfs using near-ultraviolet observations. We showed that the magnetic activity evolution differs across spectral types and is also influenced by stellar metallicity. Building on that work, this paper combines the sample from Paper \ref{cite:paper1} with LAMOST spectra to conduct a systematic investigation of the evolution of \cahk emissions in late-type stars. The paper is organized as follows. Section \ref{sec:method} describes our sample selection and the computation of chromospheric activity indices. In Section \ref{sec:result} we present our results. Finally, a discussion and a summary are given in Sections \ref{sec:discussion} and \ref{sum.sec}, respectively.

\begin{table*}
\begin{center}
{\scriptsize
  \setlength{\tabcolsep}{1pt}
\caption{Stellar parameters, stellar activity indices, stellar ages of open cluster members.}
\label{tab:table1}
\begin{tabular}{lcccccccccccc}
\toprule
ID & R.A. & Decl. & $T_{\rm{eff}}$ & log$g$ & [Fe/H] & Distance & $(B-V)_{0}$ & $(BP-RP)_{0}$ & $S_{\rm{MW}}$ & log$_{10}$(Age) & $\text{log}_{10}(R_{\text{HK}}^{'})$ & S/N \\
 & degree & degree & (K) & (dex) & (dex) & (pc) & (mag) & (mag) &  &  &  &  \\
 (1) & (2) & (3) & (4) & (5) & (6) & (7) & (8) & (9) & (10) & (11) & (12) & (13) \\ 
\hline
\midrule
43575746050631808 & 58.19351 & 17.10035 & 6040$\pm$66 & 4.19$\pm$0.1 & $-$0.13$\pm$0.07 & 985$\pm$36 & 0.55 & 0.72 & 0.34$\pm$0.15 & 8.93 & $-$4.38$\pm$0.27 & 24.58 \\
45586099981415424 & 61.57953 & 16.61004 & 6155$\pm$60 & 4.23$\pm$0.09 & $-$0.14$\pm$0.06 & 1426$\pm$63 & 0.53 & 0.76 & 0.48$\pm$0.18 & 8.93 & $-$4.17$\pm$0.2 & 23.87 \\
46167363676264320 & 60.42560 & 15.89299 & 5082$\pm$31 & 3.27$\pm$0.04 & $-$0.71$\pm$0.03 & 920$\pm$20 & 0.88 & 0.99 & 0.22$\pm$0.05 & 8.93 & $-$4.89$\pm$0.15 & 83.96 \\
46230585594749824 & 60.82205 & 16.56147 & 6136$\pm$33 & 4.17$\pm$0.05 & 0.11$\pm$0.03 & 1060$\pm$32 & 0.53 & 0.79 & 0.18$\pm$0.08 & 8.93 & $-$4.81$\pm$0.38 & 48.22 \\
46347782365527936 & 61.64106 & 16.86522 & 5936$\pm$48 & 4.29$\pm$0.07 & $-$0.25$\pm$0.05 & 762$\pm$16 & 0.58 & 0.76 & 0.23$\pm$0.07 & 8.93 & $-$4.65$\pm$0.23 & 36.38 \\
46471516080239232 & 59.75991 & 16.75634 & 4864$\pm$60 & 4.64$\pm$0.09 & $-$0.33$\pm$0.06 & 1034$\pm$45 & 0.98 & 1.11 & 0.19$\pm$0.27 & 8.93 & $-$5.04$\pm$0.71 & 24.68 \\
46566761274958848 & 59.60302 & 17.16134 & 6178$\pm$28 & 4.22$\pm$0.04 & $-$0.21$\pm$0.02 & 824$\pm$13 & 0.52 & 0.62 & 0.17$\pm$0.03 & 8.93 & $-$4.85$\pm$0.15 & 130.38 \\
46663690097339392 & 60.29038 & 17.47233 & 6044$\pm$73 & 4.28$\pm$0.11 & $-$0.34$\pm$0.07 & 1096$\pm$41 & 0.55 & 0.76 & 0.29$\pm$0.25 & 8.93 & $-$4.47$\pm$0.54 & 26.39 \\
47804394753757056 & 64.05469 & 18.88450 & 3950$\pm$41 & 4.62$\pm$0.08 & 0.11$\pm$0.09 & 46$\pm$1 & 1.39 & 1.85 & 2.31$\pm$0.08 & 8.90 & $-$4.55$\pm$0.02 & 135.35 \\
48167955146062720 & 65.79015 & 19.20374 & 4117$\pm$35 & 4.61$\pm$0.05 & $-$0.09$\pm$0.03 & 121$\pm$1 & 1.30 & 1.59 & 1.39$\pm$0.15 & 8.10 & $-$4.64$\pm$0.05 & 50.33 \\
... & ... & ... & ... & ... & ... & ... & ... & ... & ... & ... & ... & ... \\

\bottomrule
\end{tabular}}
(This table is available in its entirety in machine-readable form in the on line article.)
\end{center}
\end{table*}

\begin{table*}
\begin{center}
{\scriptsize
  \setlength{\tabcolsep}{1pt}
\caption{Stellar parameters, stellar activity indices, stellar ages of field stars.}
\label{tab:table2}
\begin{tabular}{lcccccccccccc}
\toprule
ID & R.A. & Decl. & $T_{\rm{eff}}$ & log$g$ & [Fe/H] & Distance & $(B-V)_{0}$ & $(BP-RP)_{0}$ & $S_{\rm{MW}}$ & log$_{10}$(Age) & $\text{log}_{10}(R_{\text{HK}}^{'})$ & S/N \\
 & degree & degree & (K) & (dex) & (dex) & (pc) & (mag) & (mag) &  &  &  &  \\
 (1) & (2) & (3) & (4) & (5) & (6) & (7) & (8) & (9) & (10) & (11) & (12) & (13) \\ 
\hline
\midrule
38655544960 & 45.00492 & 0.01991 & 4676$\pm$31 & 4.67$\pm$0.04 & $-$0.18$\pm$0.03 & 314$\pm$2 & 1.07 & 1.21 & 0.52$\pm$0.1 & 9.59 & $-$4.69$\pm$0.09 & 48.28 \\
549755818112 & 45.04827 & 0.04831 & 4960$\pm$31 & 4.61$\pm$0.04 & $-$0.26$\pm$0.03 & 615$\pm$11 & 0.93 & 1.13 & 0.33$\pm$0.09 & 9.82 & $-$4.71$\pm$0.13 & 61.00 \\
1275606125952 & 44.99326 & 0.07638 & 5452$\pm$74 & 4.46$\pm$0.12 & $-$0.10$\pm$0.08 & 1476$\pm$110 & 0.75 & 0.9 & 0.27$\pm$0.26 & 9.87 & $-$4.64$\pm$0.61 & 20.66 \\
2920577765120 & 45.16461 & 0.15452 & 5810$\pm$69 & 4.29$\pm$0.11 & $-$0.25$\pm$0.07 & 1563$\pm$86 & 0.63 & 0.8 & 0.11$\pm$0.2 & 9.82 & $-$5.52$\pm$4.19 & 24.63 \\
4230543624320 & 45.06346 & 0.19567 & 5971$\pm$67 & 4.22$\pm$0.11 & 0.18$\pm$0.07 & 1986$\pm$141 & 0.57 & 0.79 & 1.06$\pm$0.53 & 9.35 & $-$3.79$\pm$0.24 & 10.02 \\
6085969468928 & 44.92562 & 0.16888 & 5528$\pm$81 & 4.40$\pm$0.13 & 0.14$\pm$0.09 & 1816$\pm$155 & 0.71 & 0.88 & 0.09$\pm$0.20 & 9.83 & $-$6.07$\pm$11.67 & 21.82 \\
6223408420864 & 44.93745 & 0.18851 & 4783$\pm$37 & 4.62$\pm$0.05 & $-$0.41$\pm$0.03 & 318$\pm$2 & 1.02 & 1.11 & 0.28$\pm$0.17 & 9.92 & $-$4.91$\pm$0.30 & 42.56 \\
6944962925824 & 44.97581 & 0.18788 & 5667$\pm$30 & 4.17$\pm$0.04 & 0.20$\pm$0.02 & 970$\pm$22 & 0.68 & 0.79 & 0.14$\pm$0.03 & 9.88 & $-$5.14$\pm$0.28 & 101.20 \\
8044474553216 & 44.92765 & 0.21918 & 5042$\pm$35 & 4.64$\pm$0.05 & 0.12$\pm$0.03 & 591$\pm$11 & 0.89 & 1.08 & 0.25$\pm$0.21 & 9.66 & $-$4.82$\pm$0.44 & 38.35 \\
9281425163264 & 45.16498 & 0.20028 & 6105$\pm$15 & 4.14$\pm$0.02 & 0.11$\pm$0.01 & 261$\pm$1 & 0.54 & 0.65 & 0.20$\pm$0.02 & 9.50 & $-$4.74$\pm$0.09 & 210.62 \\
... & ... & ... & ... & ... & ... & ... & ... & ... & ... & ... & ... & ... \\

\bottomrule
\end{tabular}}\\
(This table is available in its entirety in machine-readable form in the on line article.)
\end{center}
\end{table*}

\section{Sample and Methods}
\label{sec:method}
\subsection{Sample construction}
\subsubsection{Open Clusters}
Members of open clusters exhibit similar stellar properties, making them well-suited for age determination via isochrone fitting. In Paper \ref{cite:paper1}, we constructed a large sample of open cluster members with 1,573,928 targets, incorporating data from \cite{2020A&A...640A...1C, 2020AJ....160..279K, 2022A&A...660A...4H, 2022A&A...661A.118C, 2023ApJS..264....8H, 2024A&A...686A..42H}. For a detailed description of this sample selection, we refer to Section 2.1 of Paper \ref{cite:paper1}. This catalog was then cross-matched with the LAMOST Data Release 12 (DR12) to obtain low-resolution spectra.

\subsubsection{Field stars}
As established by previous studies, stellar chemical abundances serve as reliable indicators of stellar age \citep{2013A&A...560A.109H, 2017ApJS..232....2X}. Recently, \cite{2025ApJS..280...13W} constructed a training set of stars with well-determined ages, including wide binaries, field stars, and open cluster members, and applied the XGBoost machine learning algorithm to LAMOST spectra to derive ages for around 4 million late-type dwarfs. This extensive sample provides an excellent opportunity to study the evolution of chromospheric activity in late-type stars. For our work, we built a sample of field stars by cross-matching low-resolution spectra from LAMOST DR12 with this age catalog.

\subsubsection{Stellar parameters}
In this study, we retained only spectra with a signal-to-noise ratio (S/N) greater than 10 in the \emph{g} band. Stellar parameters, including effective temperature ($T_{\rm{eff}}$), surface gravity (log$g$), and metallicity ([Fe/H]), were collected from LAMOST DR12 catalog, in which they were derived using the LAMOST Stellar Parameter Pipeline \citep[LASP;][]{2015RAA....15.1095L}. For stars with multiple observations, we selected the spectrum with the highest S/N. Meanwhile, we applied a cut of log$g \geq 3.5$ to both the open cluster members and field stars to only keep dwarfs in our sample. Additionally, \emph{Gaia} magnitudes, \emph{renormalized unit weight error} (ruwe) values and distances were obtained from the Gaia early Data Release 3 \citep[eDR3;][]{2021A&A...649A...1G} and the catalog of \cite{2021AJ....161..147B}, respectively. To ensure reliable distance measurements, we retained only stars located within 5 kpc and with relative parallax errors smaller than 0.2. 
Finally, to minimize contamination from binary systems, we excluded targets with ruwe exceeding 1.4. 

The color excess values $E(B-V)$ were obtained from \cite{2019ApJ...887...93G} and \cite{1998ApJ...500..525S}, while interstellar extinction was calculated using the coefficients from \cite{1999PASP..111...63F}. In Paper \ref{cite:paper1}, we categorized the sample into different subtypes based on their $(BP-RP)_{0}$ colors, then for each subtype we derived two typical ages to mark its entering and existing points of the main sequence with the MESA Isochrones and Stellar Tracks \citep{2011ApJS..192....3P, 2016ApJ...823..102C, 2016ApJS..222....8D}. Following a similar approach, in this work we divided our sample into four types (F/G/K/M) according to their $(BP-RP)_{0}$ colors and excluded stars that have already left the main sequence by comparing their measured ages against the typical age values.

\subsection{S-index and $R_{\rm{HK}}^{'}$}
\label{sec:index}
One of the most widely used proxies for chromospheric activity is the \cahk emission, which at first were usually quantified using the well-known $S-$index based on the Mount Wilson observations ($S_{\rm{MW}}$) \citep{1978PASP...90..267V}. It was defined as the ratio of fluxes between the emission cores of \cahk lines and the continua at both sides of the lines. However, this index still contains contribution from the photosphere. Later works then converted the $S-$index into pure chromospheric activity index $R_{\rm{HK}}^{'}$ \citep{1982A&A...107...31M, 1984ApJ...279..763N}. 

It should be noted that there are multiple approaches to derive $R_{\rm{HK}}^{'}$ from spectroscopic data \citep[e.g.,][]{2013A&A...549A.117M,2017A&A...600A..13A,2018A&A...616A.108B}. Recent work by \cite{2021A&A...652A.116P} showed that during $R_{\rm{HK}}^{'}$ calculation, direct \cahk flux measurement (without S-index conversion) from observed spectra and using the synthetic spectra for photospheric rectification may yield more reliable results when the spectral resolution is high than $\approx$20,000. Given that LAMOST spectra have a resolution of only R$\sim$1800, we used the classical method to convert $S$-index to $R_{\rm{HK}}^{'}$. We calculated the $S-$indices of LAMOST spectra ($S_{\rm{LAMOST}}$) following the definition of \cite{1978PASP...90..267V}. 

Furthermore, to convert the $S_{\rm{LAMOST}}$ into the pure chromospheric activity index $R_{\rm{HK}}^{'}$, an additional calibration between $S_{\rm{LAMOST}}$ and Mount Wilson $S_{\rm{MW}}$ is necessary. We adopted the calibration established by \cite{2025ApJ...984....2H}, which is valid for stars with $\rm{[Fe/H]} > -1$. In brief, the authors simulated chromospheric activity by adding gaussian profiles to PHOENIX synthetic spectra \citep{2013A&A...553A...6H}, which were further convolved to different resolutions. Later on, relations between $S_{\rm SW}$ and $S_{\rm SPECTRUM}$ (e.g., $S_{\rm LAMOST}$) corresponding to different resolutions have been established. Such conversion could minimize the influence of low spectral resolution. A detailed description of the calculation 
processes are given in \citet{2025ApJ...984....2H}. We adopted the methods of \cite{1984ApJ...279..763N} and \cite{1982A&A...107...31M} to calculate $R_{\rm{HK}}^{'}$:
\begin{equation}
    R_{\rm{HK}}^{'} = 1.34 \times 10^{-4} \times C_{\rm cf}^{'} \times S_{\rm{MW}} - R_{\rm{phot}}.
    \label{rhk.eq}
\end{equation}
Here $R_{\rm{phot}}$ is the photospheric contribution and $C_{\rm{cf}}^{'}$ is the conversion factor. Both are functions of the stellar intrinsic color $(B-V)_{0}$, which were calculated by interpolating the $T_{\rm{eff}}$ using the results from \cite{2013ApJS..208....9P}.

\begin{figure}
\centering
\includegraphics[width=0.45\textwidth]{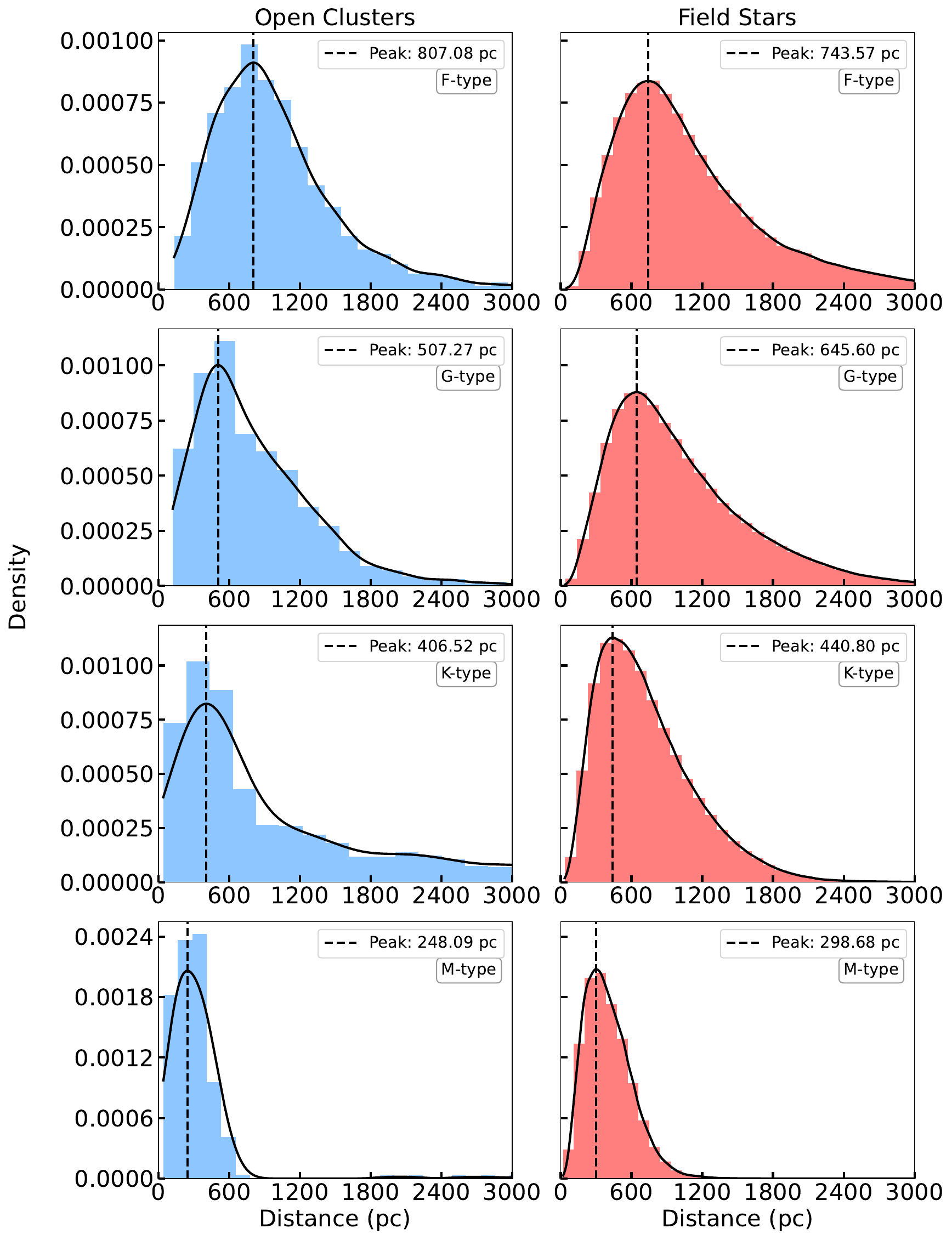}
\caption{Histograms of distances corresponding to open cluster members and field stars across various stellar types. Peak positions are marked using black dashed lines.}
\label{dist.fig}
\end{figure}

\begin{figure*}
\centering
\subfigure[]{
\includegraphics[width=0.47\textwidth]{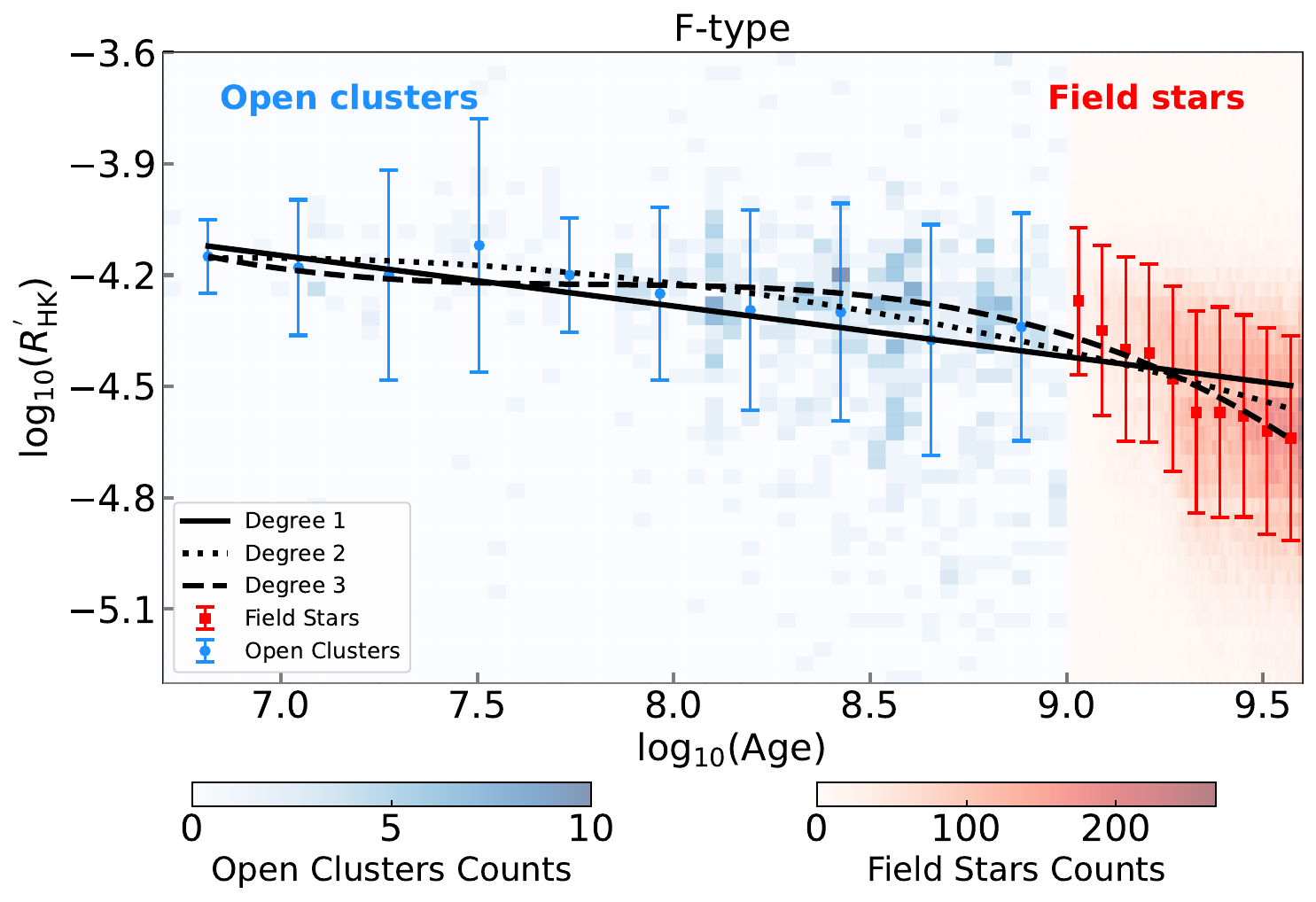}}
\subfigure[]{
\includegraphics[width=0.47\textwidth]{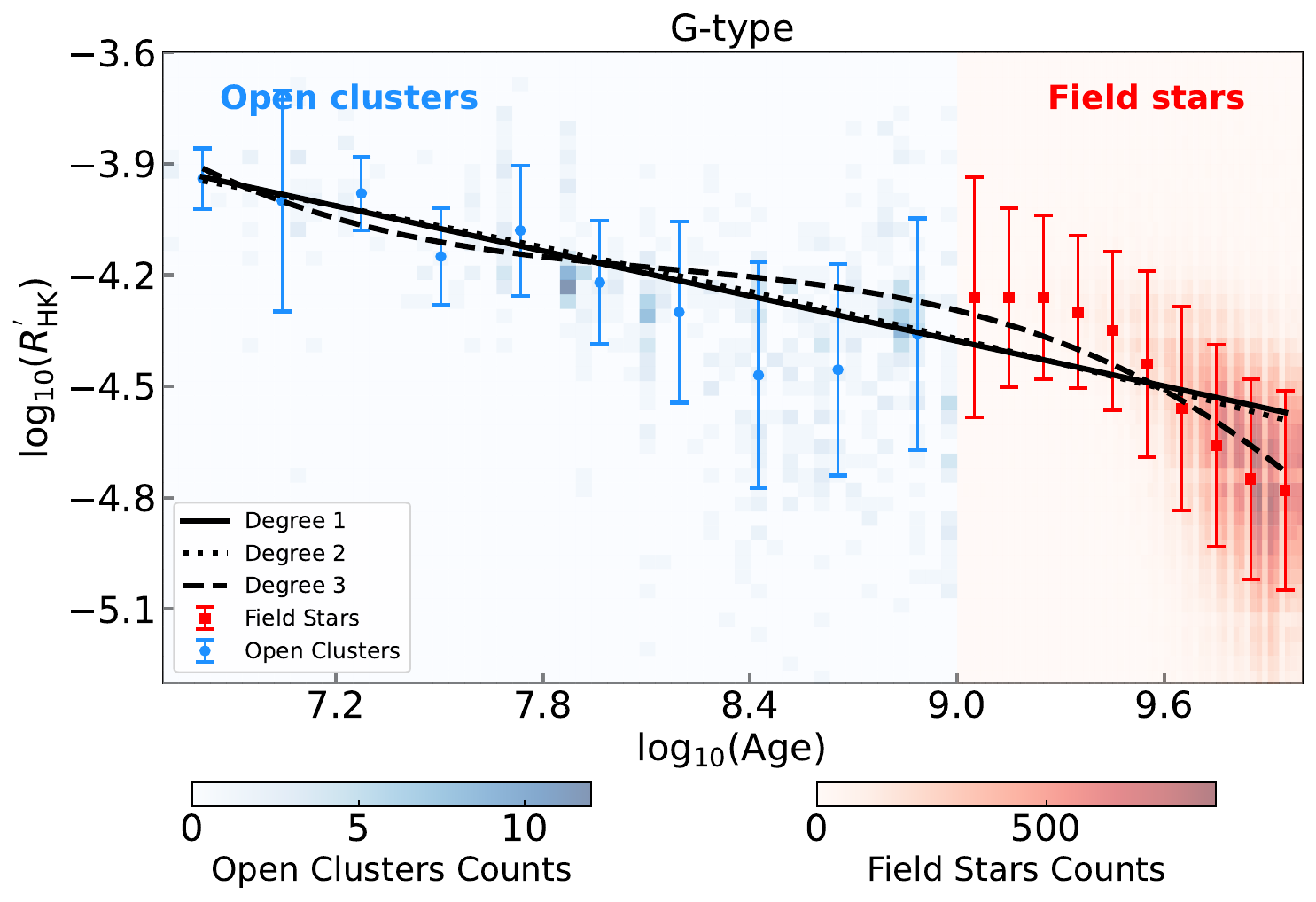}}
\subfigure[]{
\includegraphics[width=0.47\textwidth]{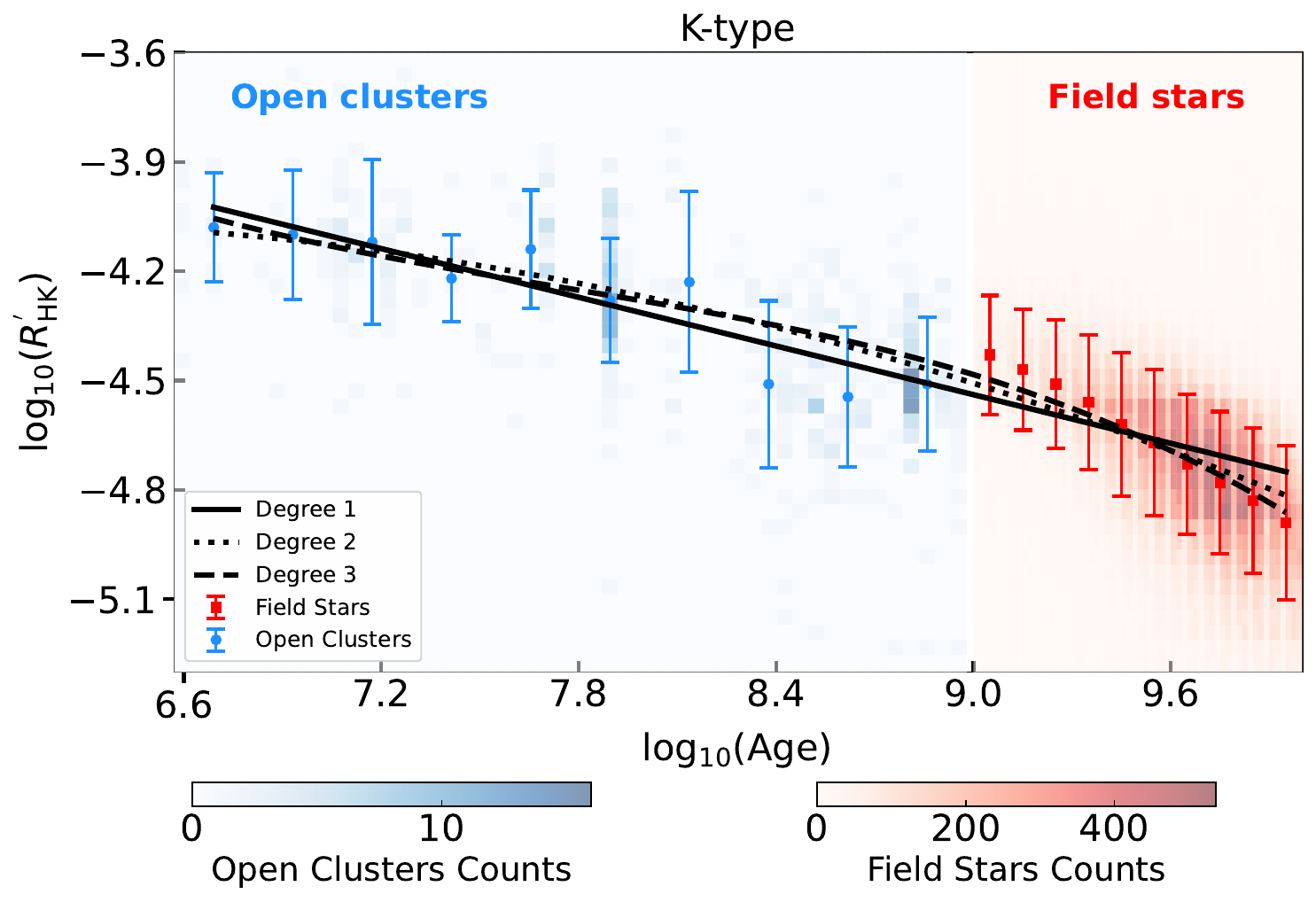}}
\subfigure[]{
\includegraphics[width=0.47\textwidth]{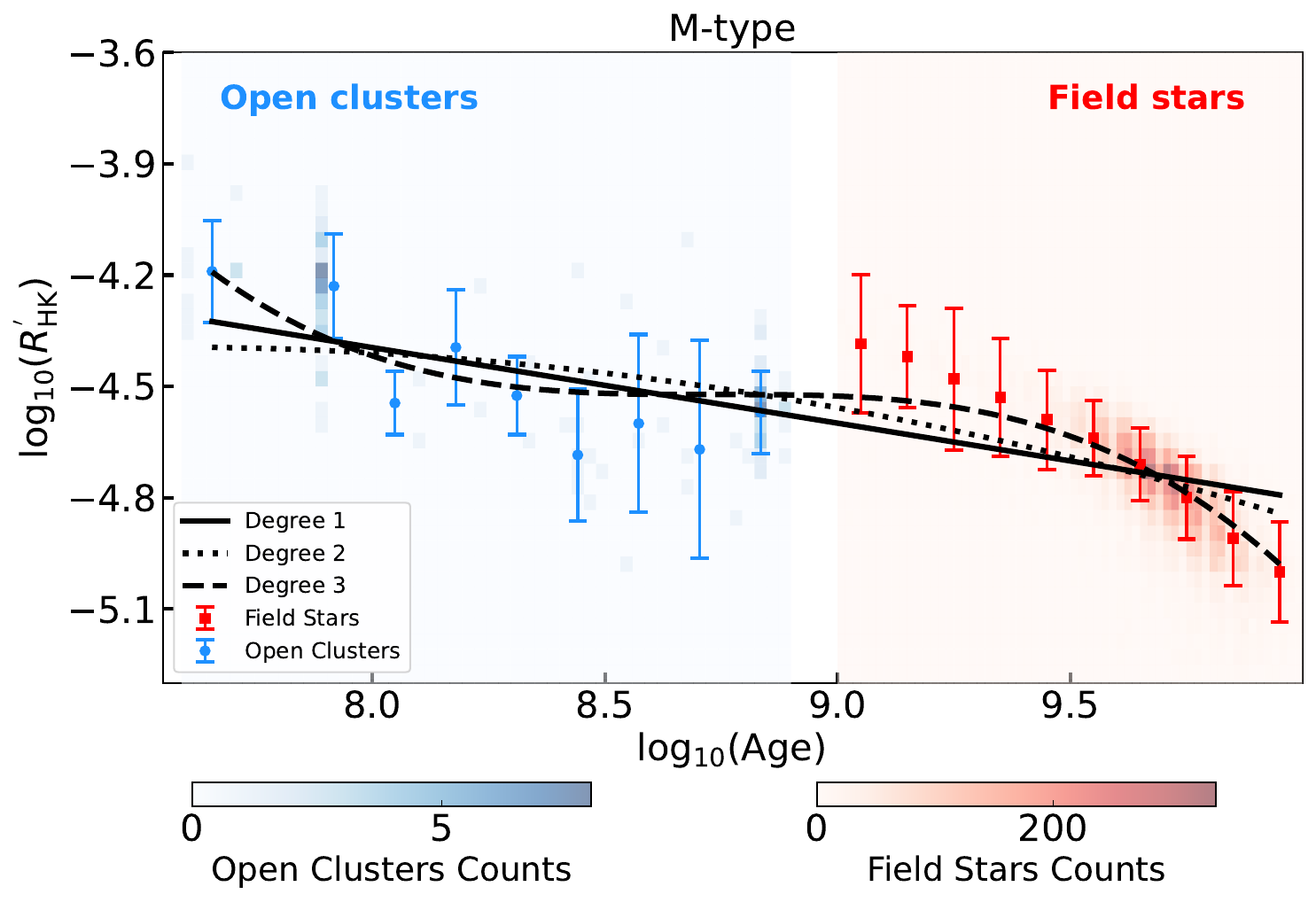}}
\caption{$R_{\rm{HK}}^{'}$--age relations of different kinds of stars. Panel (a), (b), (c), and (d) represents F-, G-, K-, and M-type dwarfs, respectively. Members of open clusters are presented in blue while fields stars are presented in red. Blue and red dots with errorbars are binned median points. Black lines are best-fit models corresponding to maximum a posteriori.}
\label{relation.fig}
\end{figure*}

\section{Results}
\label{sec:result}

In this work, we combined both open cluster members and field stars to investigate the evolution of chromospheric activity across different evolutionary stages.
We listed all our targets together with stellar parameters, activity indices and ages in Table \ref{tab:table1} and Table \ref{tab:table2}. In total we got 5,748 open cluster members and 1,674,191 field stars. 
We divided our sample into subgroups according to their $T_{\rm{eff}}$. For F-type, G-type, K-type, and M-type dwarfs we adopted $6,000 \, \text{K} \leq T_{\rm{eff}} < 6,500 \, \text{K}$, $5,300 \, \text{K} < T_{\rm{eff}} \leq 6,000 \, \text{K}$, $4,000 \, \text{K} < T_{\rm{eff}} \leq 5,300 \, \text{K}$, and $3,800 \, \text{K} < T_{\rm{eff}} \leq 4,000 \, \text{K}$, respectively. For each subgroup, we retained only targets with distances smaller than the peak of the distance histogram specific to that subgroup, ensuring a volume-complete sample. This completeness cut was applied separately to the open clusters and field stars (Figure \ref{dist.fig}).

Following the results of \cite{2025ApJS..280...13W}, To ensure reliable age estimation, we restricted our analysis of field stars to those with spectra having a S/N greater than 20 and ages between 1 and 10 Gyr. In contrast, for the open cluster sample, given the limited number of sample older than 1 Gyr, we only focused on target younger than 1 Gyr. To ensure reliable measurements of $R_{\rm{HK}}^{'}$, we further limited our sample to stars with activity levels in the range $10^{-5.5} \leq R_{\rm{HK}}^{'} \leq 10^{-3}$. 
Meanwhile, for each spectral subtype we binned the data along the age axis and computed the median trend (Figure \ref{relation.fig}). For both open cluster members and field stars, the observed age range was divided into ten bins in the log space. Within each bin, the median
log$_{10}(R_{\rm{HK}}^{'})$ value was taken as the representative activity level, and the associated uncertainty was estimated from the scatter of the data (i.e., standard deviation) in that bin.

\begin{table*}
\centering
\caption{Best-fit parameters and their asymmetric uncertainties for various models.}
\label{tab:table3}
\setlength{\tabcolsep}{12pt}
\begin{tabular}{l l r r r r r r r}
\toprule
Type & Degree & $a_3$ & $a_2$ & $a_1$ & $a_0$ & $\Delta\ln Z_{3}$ & $\Delta\ln Z_{2}$ \\
\midrule

F & 3rd & 32.406864 & $-$14.106963 & 1.8111 & $-$0.077505 & - & - \\
  &     & $( {}^{+7.810729}_{-32.33927} )$ & $( {}^{+12.545303}_{-2.483614} )$ & $( {}^{+0.263676}_{-1.61206} )$ & $( {}^{+0.068663}_{-0.009086} )$ &  &  \\
  & 2nd & 0 & $-$6.895304 & 0.793518 & $-$0.057422 & 8.26$\pm$0.21 & - \\
  &     & - & $( {}^{+4.85691}_{-0.555807} )$ & $( {}^{+0.117668}_{-1.206769} )$ & $( {}^{+0.074243}_{-0.005558} )$ &  &  \\
  & 1st & 0 & 0 & $-$3.192104 & $-$0.136439 & 16.28$\pm$0.19 & 8.02$\pm$0.18 \\
  &     & - & - & $( {}^{+0.385153}_{-0.376219} )$ & $( {}^{+0.045864}_{-0.047042} )$ &  &  \\
\midrule

G & 3rd & 30.672866 & $-$12.65184 & 1.541196 & $-$0.063014 & - & - \\
  &     & $( {}^{+16.415716}_{-26.186224} )$ & $( {}^{+9.733918}_{-5.997482} )$ & $( {}^{+0.725148}_{-1.191598} )$ & $( {}^{+0.048519}_{-0.029119} )$ &  &  \\
  & 2nd & 0 & $-$3.300006 & $-$0.018874 & $-$0.011136 & 7.73$\pm$0.21 & - \\
  &     & - & $( {}^{+2.750964}_{-2.500602} )$ & $( {}^{+0.614954}_{-0.675009} )$ & $( {}^{+0.041134}_{-0.037133} )$ &  &  \\
  & 1st & 0 & 0 & $-$2.556173 & $-$0.202412 & 15.5$\pm$0.19 & 7.77$\pm$0.18 \\
  &     & - & - & $( {}^{+0.300929}_{-0.316545} )$ & $( {}^{+0.039553}_{-0.037614} )$ &  &  \\
\midrule

K & 3rd & 11.88574 & $-$5.912632 & 0.738307 & $-$0.031492 & - & - \\
  &     & $( {}^{+26.852062}_{-13.338222} )$ & $( {}^{+5.012869}_{-9.749875} )$ & $( {}^{+1.173542}_{-0.625596} )$ & $( {}^{+0.025862}_{-0.046964} )$ &  &  \\
  & 2nd & 0 & $-$5.70103 & 0.550257 & $-$0.04641 & 7.46$\pm$0.2 & - \\
  &     & - & $( {}^{+3.986852}_{-0.73011} )$ & $( {}^{+0.184412}_{-0.973174} )$ & $( {}^{+0.058596}_{-0.011497} )$ &  &  \\
  & 1st & 0 & 0 & $-$2.535415 & $-$0.222606 & 15.75$\pm$0.19 & 8.29$\pm$0.18 \\
  &     & - & - & $( {}^{+0.322496}_{-0.33898} )$ & $( {}^{+0.039595}_{-0.037625} )$ &  &  \\
\midrule

M & 3rd & 171.556481 & $-$60.482639 & 6.925193 & $-$0.264304 & - & - \\
  &     & $( {}^{+10.846179}_{-78.643357} )$ & $( {}^{+26.873159}_{-3.393348} )$ & $( {}^{+0.349516}_{-3.055421} )$ & $( {}^{+0.115555}_{-0.011793} )$ &  &  \\
  & 2nd & 0 & $-$8.83914 & 1.177009 & $-$0.077924 & 3.37$\pm$0.21 & - \\
  &     & - & $( {}^{+7.515438}_{-0.041887} )$ & $( {}^{+0.021025}_{-1.712276} )$ & $( {}^{+0.096533}_{-0.001855} )$ &  &  \\
  & 1st & 0 & 0 & $-$2.765918 & $-$0.203693 & 11.25$\pm$0.2 & 7.88$\pm$0.18 \\
  &     & - & - & $( {}^{+0.362286}_{-0.364146} )$ & $( {}^{+0.040981}_{-0.040461} )$ &  &  \\
\bottomrule
\end{tabular}

\smallskip
\footnotesize
Note: $\Delta \ln Z_{j} = \ln Z_{\text{current row model}} - \ln Z_{\text{model } j}$. Positive values indicate the current model is favored over model $j$, where $j$ denotes the polynomial degree. Uncertainties for coefficients are shown in parentheses in the row below each model.
\end{table*}

Besides the classical Skumanich-type relation, polynomial with different degrees have  been carried out to model the evolution of stellar activity \citep[e.g.][]{2008ApJ...687.1264M, 2019ApJ...887...84Z}. In this work, we applied 1st-, 2nd-, and 3-rd degree polynomial:
\begin{equation}
    \log_{10}(R'_{\rm{HK}}) = \sum_{i=0}^{n} a_i \, (\text{log}_{10}t)^{\,n - i}, \quad n = 1, 2, 3
\end{equation}
to fit the $R_{\rm{HK}}^{'}-$age relation. Here $t$ is stellar age.

We performed Bayesian model comparison using the \emph{dynesty} package \citep{2020MNRAS.493.3132S, sergey_koposov_2025_17268284}, which implements nested sampling \citep{2004AIPC..735..395S, 10.1214/06-BA127}. We used dynamic nested sampling \citep{2019S&C....29..891H} with multi-ellipsoid bounding \citep{2009MNRAS.398.1601F} and random walk \citep{10.1214/06-BA127}. The maximum a posteriori (MAP) estimation was obtained by selecting the sample with the highest likelihood from the resampled posterior distribution, which we adopted as our best-fit solution. The uncertainties of the \emph{evidence} were estimated following \cite{2008arXiv0801.3887C, 10.1214/17-BA1075, 2020MNRAS.493.3132S}. Uniform priors were applied to all model parameters. The likelihood function was written as: 
\begin{equation}
\begin{aligned}
    \ln\big[p(y \mid x, \boldsymbol{\sigma}, \mathbf{a})\big] 
    = -\frac{1}{2} \sum_{n} \frac{\left[ y_n - \sum_{i=0}^{n_{\text{deg}}} a_i \, x_n^{\,n_{\text{deg}} - i} \right]^2}{\sigma_n^2} \\
      - \frac{1}{2} \sum_{n} \ln\!\big(2\pi \sigma_n^2\big),
\end{aligned}
\end{equation}
Here \emph{y} is observed chromsopheric activity level, \emph{x} is stellar age, and $\sigma$ is the uncertainty of $R_{\rm{HK}}^{'}$. The polynomial models were constrained to be monotonically decreasing. Examples of the resulting posterior corner plots are shown in Figure \ref{comparison.fig}. The fitting results are summarized in Table \ref{tab:table3}. Figure \ref{relation.fig} shows all the activity--age relations for different types of stars.

In addition, to test whether the error of $R_{\rm{HK}}^{'}$ in each age bin is reasonable (and has no significant impact on the fit), we applied an alternative fit for the G-type stars (as an example), in which the uncertainty of $R_{\rm{HK}}^{'}$ was set as a free parameter. 
The likelihood function was written as: 
\begin{equation}
\begin{aligned}
    \ln\big[p(y \mid x, \boldsymbol{\sigma}, \mathbf{a})\big] 
    = -\frac{1}{2} \sum_{n} \frac{\left[ y_n - \sum_{i=0}^{n_{\text{deg}}} a_i \, x_n^{\,n_{\text{deg}} - i} \right]^2}{\sigma^2} \\
      - \frac{1}{2} \sum_{n} \ln\!\big(2\pi \sigma^2\big).
\end{aligned}
\end{equation}
The fitting results (Figure \ref{comparison.fig} and \ref{comparison2.fig}) show that the derived uncertainty is smaller than the standard deviation of log$_{10}(R_{\rm{HK}}^{'})$ in each bin, yet the fitted activity--age relations remain similar. 
We therefore concluded that using the standard deviation of log$_{10}(R_{\rm{HK}}^{'}$) in each bin as the error is a robust choice for the fitting.

\begin{figure}
\centering
\includegraphics[width=0.45\textwidth]{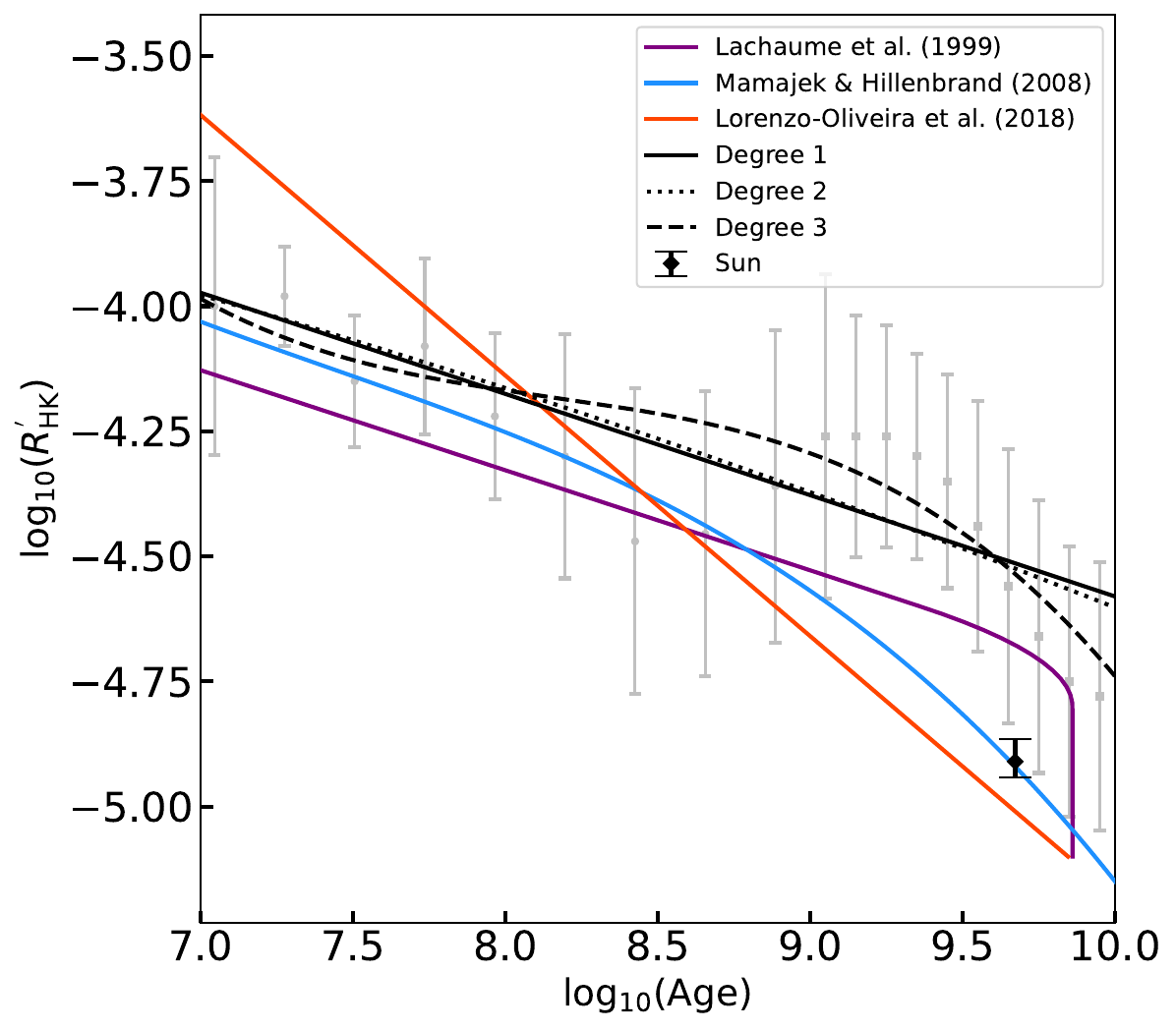}
\caption{Comparisons of $R_{\rm{HK}}^{'}-$age relations for G-type stars from different works. Gray dots with errorbars are our G-type star sample. Black lines represent results from our work with different models. Purple, dodgerblue, and orangered line shows the relation from \cite{1999A&A...348..897L}, \cite{2008ApJ...687.1264M}, and \cite{2018A&A...619A..73L}, respectively. The Sun was shown as black diamond with the $R_{\rm{HK, Sun}}^{'} = -4.91$ and its uncertainties adopted from \cite{2008ApJ...687.1264M}.}
\label{rel_comp.fig}
\end{figure}

\begin{figure*}
\centering
\includegraphics[width=0.95\textwidth]{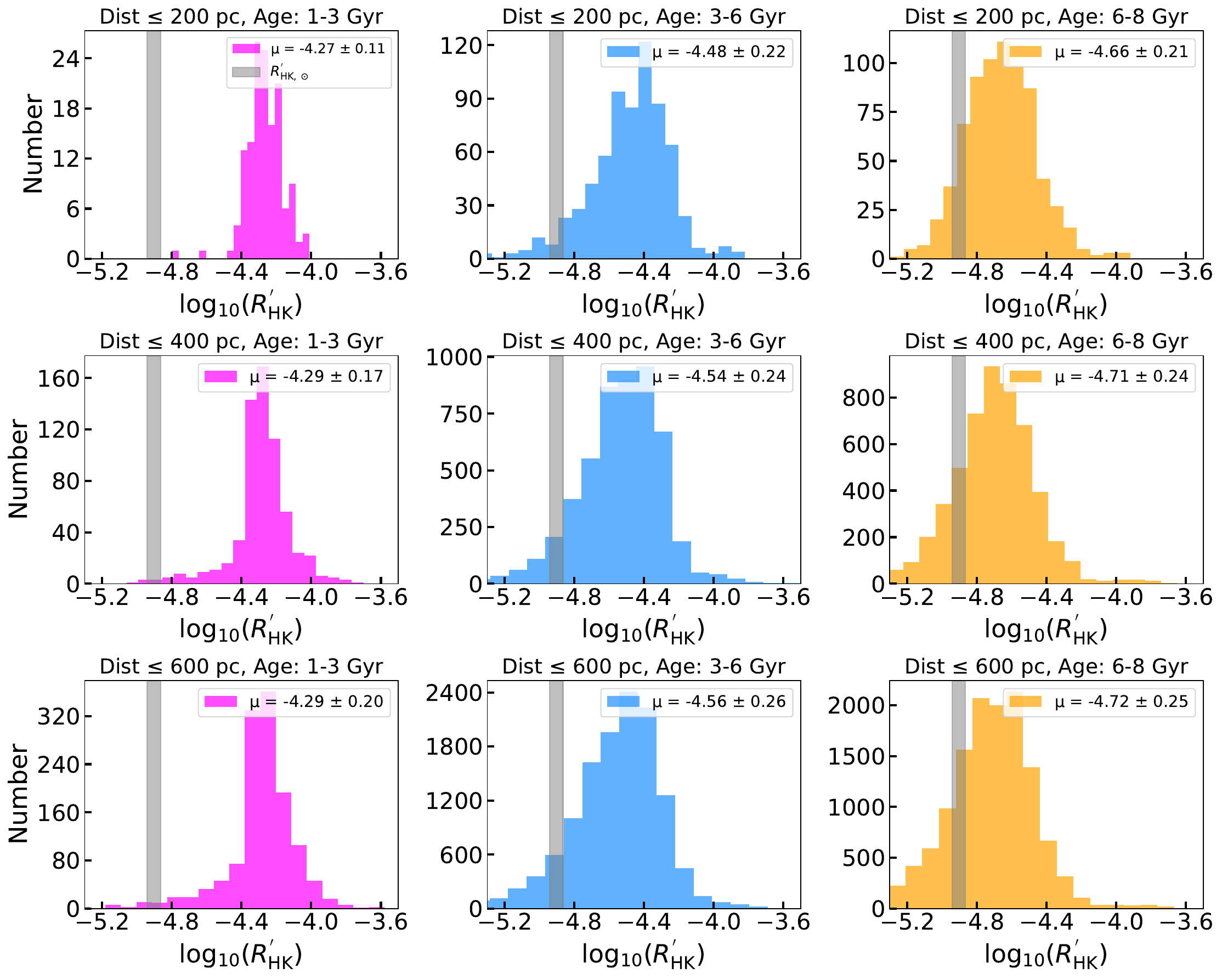}
\caption{Histograms of log$_{10}(R_{\rm{HK}}^{'})$ of solar twins. Different panels represent different age and distance ranges. Shaded areas mark the solar log$_{10}(R_{\rm{HK}}^{'})$ range adopted from \cite{2008ApJ...687.1264M}.}
\label{twin.fig}
\end{figure*}

\begin{figure*}
\centering
\subfigure[]{
\includegraphics[width=0.47\textwidth]{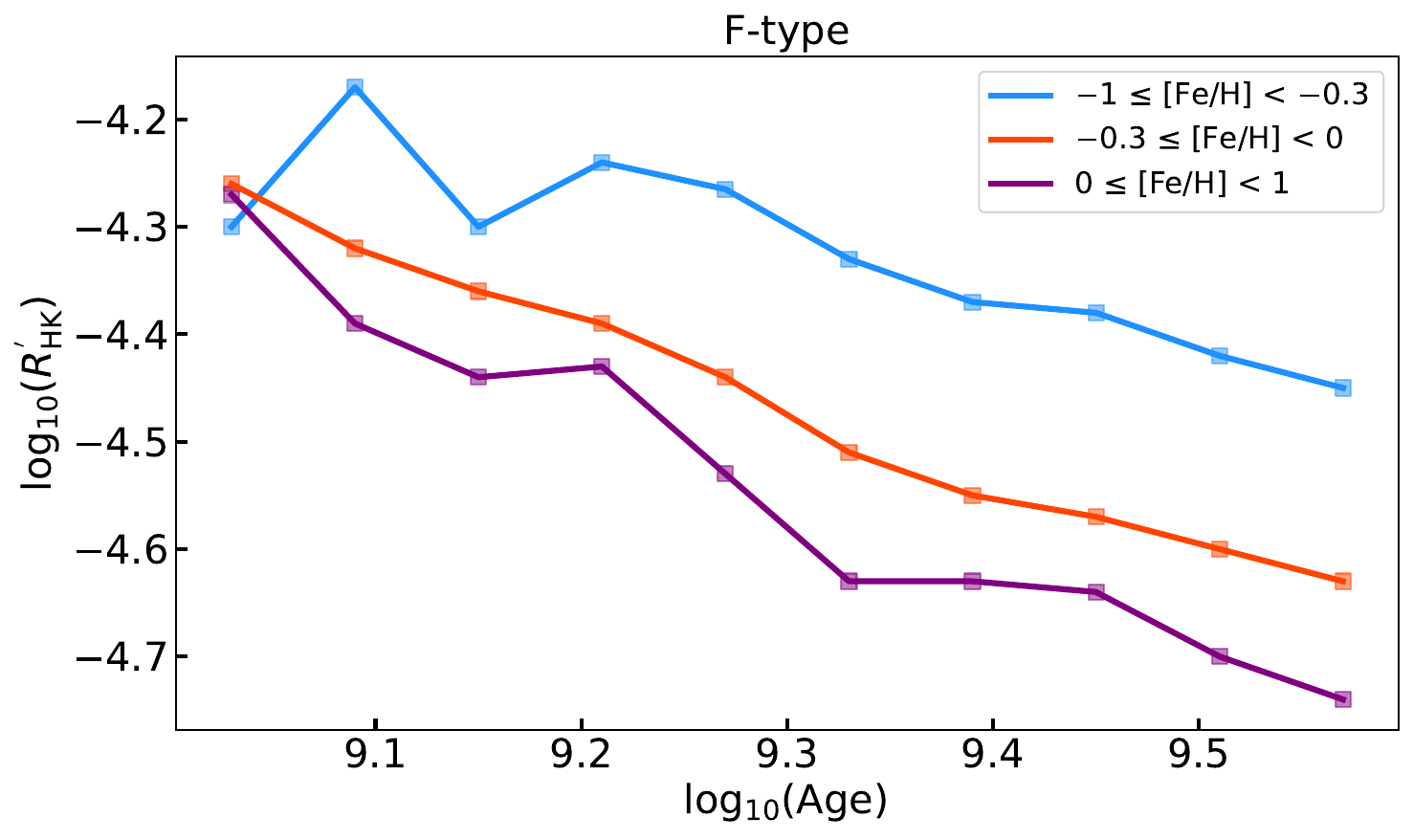}}
\subfigure[]{
\includegraphics[width=0.47\textwidth]{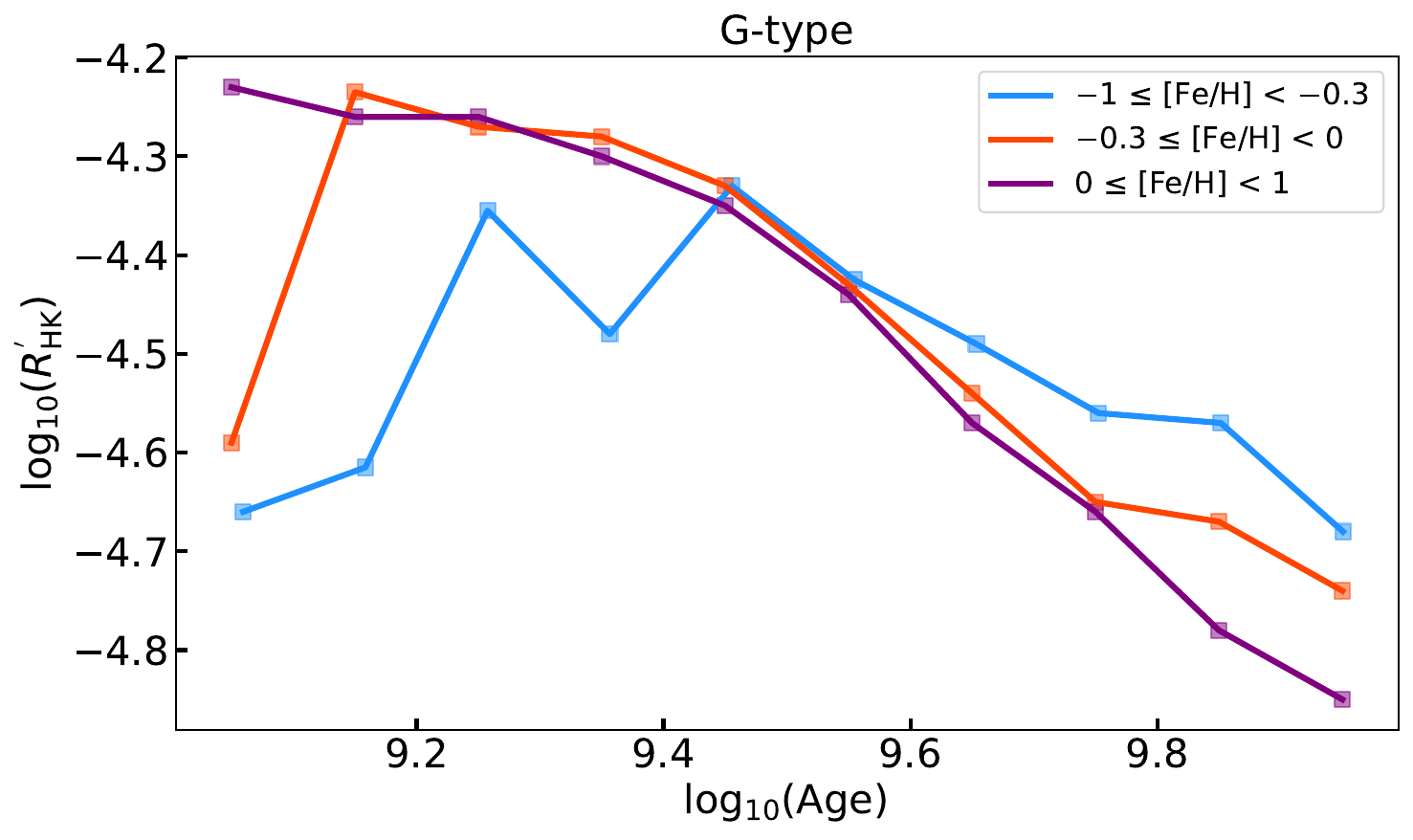}}
\subfigure[]{
\includegraphics[width=0.47\textwidth]{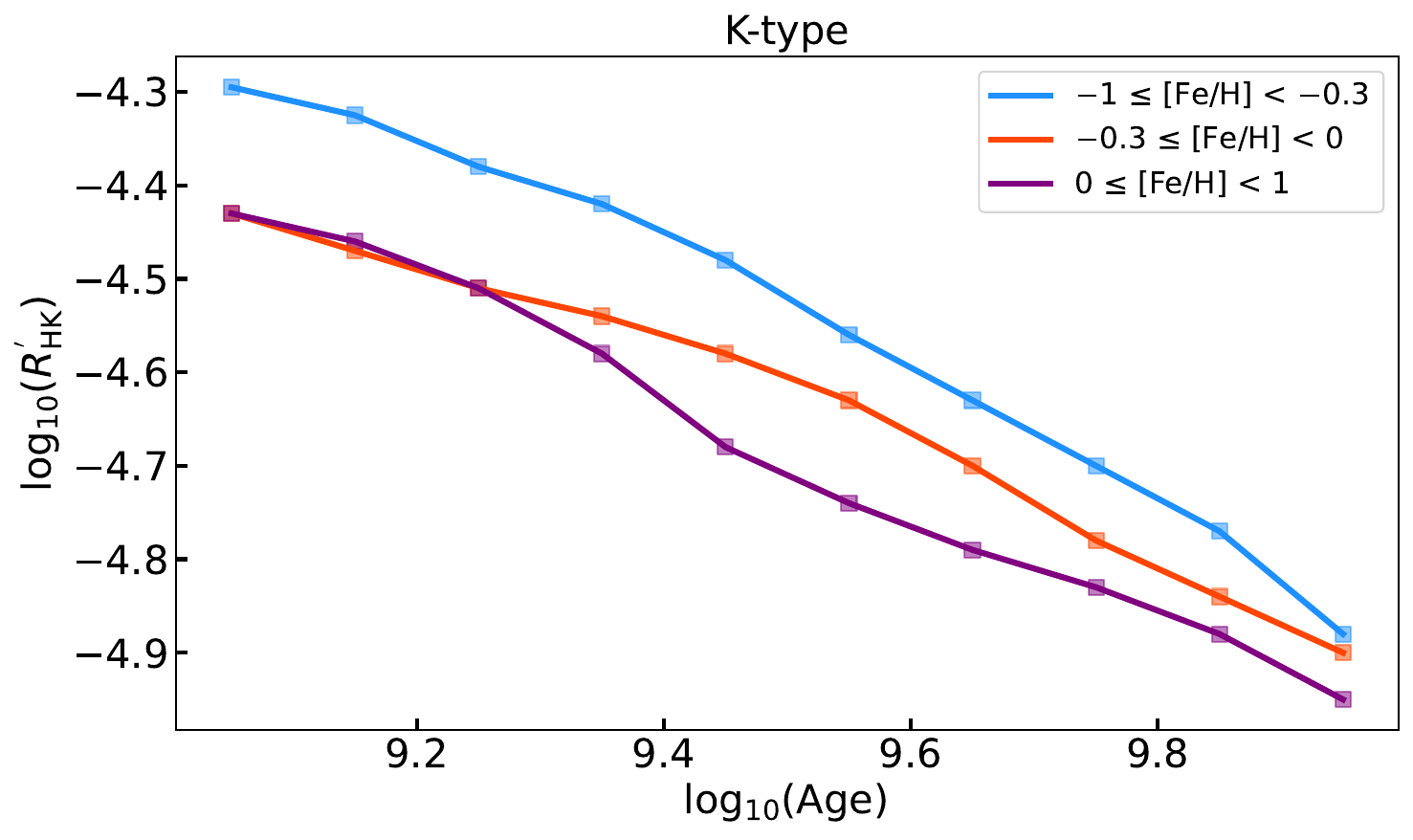}}
\subfigure[]{
\includegraphics[width=0.47\textwidth]{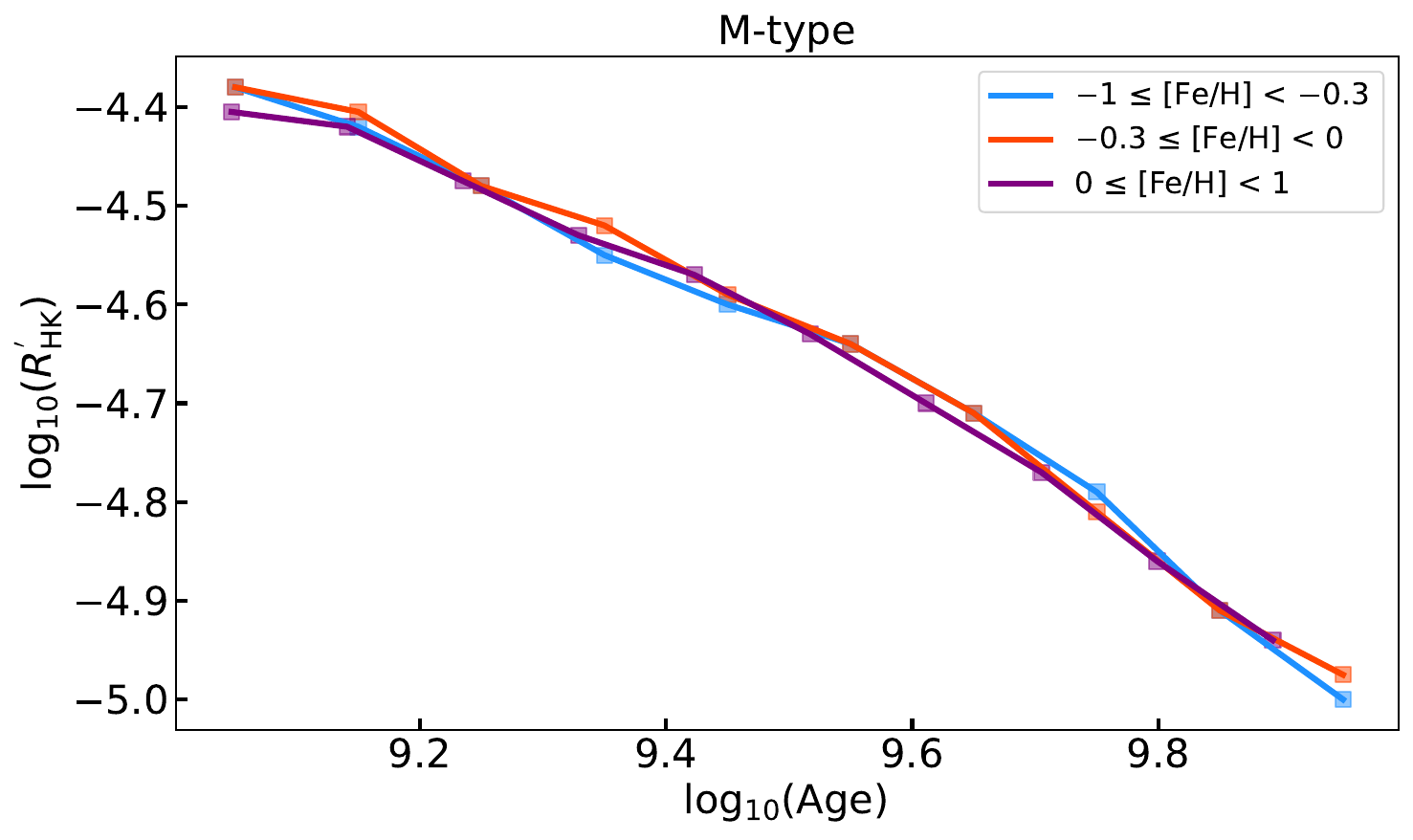}}
\caption{Median curves of $R_{\rm{HK}}^{'}$--age plane corresponding to different stellar types for field stars. Different colors represent different metallicity ranges.}
\label{metal.fig}
\end{figure*}

\section{Discussions}
\label{sec:discussion}

\subsection{Chromospheric activity--age relations}

The long-term evolution of chromospheric activity has been studied extensively. Early studies primarily focused on members of open clusters, whose ages can be reliably estimated via isochrone fitting. The classical framework was established by \cite{1972ApJ...171..565S}, who modeled this evolution using a power-law relation (i.e., a linear function in log–log space). In this paradigm, chromospheric activity levels can serve as effective age indicators. The Skumanich-law was not only observed in activity proxies but also in strength of magnetic field for both pre-main sequence stars and main sequence stars \citep{2014MNRAS.441.2361V}. 

Early studies on the evolution of chromospheric activity primarily focused on open cluster members. Several works have suggested that chromospheric activity declines rapidly and reaches a minimum on a relatively short timescale. For example, \cite{2004A&A...426.1021P} and \cite{2009A&A...499L...9P} proposed very similar timescales of approximately 1.5 Gyr and 1.2 Gyr, respectively. Later on, by incorporating field stars into the analysis, \cite{2013A&A...551L...8P} found a ``L-shape'' model for chromospheric activity evolution, featuring a knee point at around 2 Gyr, beyond which the activity level will reach its minimum and remain nearly unchanged.

However, alternative perspectives have emerged from other studies. \cite{2008ApJ...687.1264M} combined activity--rotation relations with gyrochronology to calibrate the activity--age relation for F7-K2 dwarfs aged between 0.6 and 4.5 Gyr. Later on, \cite{2018A&A...619A..73L} derived Skumanich-type activity--age relation for solar twins using HARPS spectra, with stellar ages determined through isochrone fitting. They argued that chromospheric activity remains a valid age indicator for stars up to 7 Gyr. More recent works have further highlighted the necessity of accounting for stellar mass and metallicity when constructing activity--age relations \citep[e.g.,][]{2024MNRAS.532..563S, 2025ApJ...983L..31C}.

Given that first‑, second‑, and third‑order polynomial models have all been proposed in earlier studies \citep[e.g.,][]{2008ApJ...687.1264M, 2018A&A...619A..73L, 2019ApJ...887...84Z, 2025MNRAS.542.2431L} to describe magnetic activity evolution, our large sample provides a robust opportunity to assess which of these models offers the best description of the data.
In Section \ref{sec:result}, we explored various models to describe the activity–age relation. We computed the Bayesian \emph{evidence} (ln$Z$) for each model using the \emph{dynesty} Python package \citep{2020MNRAS.493.3132S, sergey_koposov_2025_17268284} and applied Jeffreys’ scale ($\Delta\text{ln}Z$) to judge which model is better (See  \cite{2011A&A...527A..56H} and reference in it for more details). Our analysis suggests that, across all spectral types, a simple linear model best describes the activity--age relation, although with different slopes.

However, we noted that the activity--age relation exhibits some features beyond a simple linear decline. After reaching the main sequence, chromospheric activity undergoes a relatively slow decay over a substantial timescale, a trend also reported by \cite{2014AJ....148...64S} and \cite{2023ApJ...951...44R} based on X-ray and ultraviolet observations. Both studies, as well as our analysis of field stars, reveal an abrupt drop in activity around $\sim$1 Gyr. This transition is further supported by \cite{2017MNRAS.471.1012B}, who detected a similar sharp decline using X-ray data from the \emph{Chandra} and \emph{XMM-Newton} satellites.
The preference for a linear relation in this work may be driven by the large errors in the data points (Figure \ref{relation.fig}). This makes it difficult to determine whether these features are physically real, and consequently whether a higher‑order polynomial is genuinely required to describe the activity evolution.
Figure \ref{rel_comp.fig} compares our $R_{\rm{HK}}^{'}$--age relation of G dwarfs with those of previous works \citep{1999A&A...348..897L, 2008ApJ...687.1264M, 2018A&A...619A..73L}. The systematic offset in absolute $R_{\rm{HK}}^{'}$ levels likely arises from differences in the calculation methods, whereas the differing evolutionary slopes may be attributed to different stellar samples used. Given the considerable uncertainties in the data, it remains difficult to judge which model provides a more accurate description of the  activity evolution.

Recent studies have proposed a complex evolution behavior of stellar rotation. For example, core-envelope coupling \citep[e.g.][]{2013A&A...556A..36G, 2020A&A...636A..76S} and weakened magnetic braking \citep{2016Natur.529..181V} can redistribute stellar angular momentum, which will influence stellar activity levels. As a result, the relation between chromospheric activity and stellar rotation period will evolve with time. This behavior may result in non-uniform decay rate in the activity--rotation relation, which have been observed by some recent researches.

For example, \cite{2025A&A...699A.251Y} suggested a four-segment activity--rotation relation, connecting it to three phases of gyrochronology and one weakened magnetic braking stage. Later on, \cite{han2025varyingcoreenvelopecouplingefficiency} derived a new activity--rotation relation for M dwarfs and divided the intermediate region (corresponding to the ``Gap'' region of gyrochronology) into two parts with varying activity decay rate, indicating a change in the core-envelope coupling efficiency. These features suggest a varying relationship between activity and age during long-term stellar evolution. The complex behavior of activity--rotation relation is indicative of a non-uniform decay rate of chromospheric activity throughout stellar lives. More observations are needed to shed light on such issue.

The Sun exhibits log$_{10}$($R_{\rm{HK}}^{'}$) value in the range $-$4.942--$-$4.865 \citep{2008ApJ...687.1264M}, much lower than that predicted by our new relation.
To address whether the Sun is truly special, we compared its $R_{\rm{HK}}^{'}$ value with those of solar twins in our sample, defined as stars with $5,677\,K \leq T_{\rm{eff}} \leq 5,877\,K$, $4.2 \leq$ log$g$ $\leq 4.6$, and $-0.2 \leq$ [Fe/H] $\leq 0.2$ (Figure \ref{twin.fig}). The data clearly show that the chromospheric activity declines with stellar age. Although the Sun indeed lies among the least active stars, many solar twins exhibit activity levels comparable to the present-day Sun .

\subsection{Influence of metallicity}
Several studies have explored the potential influence of metallicity on $R_{\rm{HK}}^{'}$--age relations \citep[e.g.][]{2005A&A...431..329L, 2016A&A...594L...3L, 2024MNRAS.532..563S, 2025ApJ...983L..31C}. In this work, we also examined how metallicity affects chromospheric activity. Given the limited number of stars in open clusters, our analysis focuses exclusively on field stars. In Figure \ref{metal.fig}, we show median $R_{\rm{HK}}^{'}$--age trends for different spectral types, with distinct colors representing different metallicity ranges. Overall, the decay rates of activity with age are similar across different metallicity ranges, although systematic offsets in $R_{\rm{HK}}^{'}$ are evident: metal-poor stars tend to exhibit higher $R_{\rm{HK}}^{'}$ levels. Such metallicity stratification has also been revealed and calibrated by \citep{2025ApJ...983L..31C} and \cite{1998MNRAS.298..332R}.

Such phenomenon is likely due to the intrinsically shallower \cahk absorption lines for metal-poor stars, which would lead to an apparent enhancement in measured activity \citep{1991ApJ...375..722S}. We noticed that such phenomenon is not obvious for young G-type dwarfs, which may be attributed to the limited number of stars in this region. In addition, this metallicity-dependent effect disappears for M dwarfs. In these cool stars, the \cahk lines lack the broad wings seen in hotter stars because the dominant ionization stage is \cai rather than \caii that produces the broad wings \citep{2017ARA&A..55..159L}. Strong TiO molecular bands in M-dwarf spectra can obscure the shallow \cahk absorption lines, making the influence of metallicity observationally indistinguishable.

\section{Summary}
\label{sum.sec}
Although stellar chromospheric activity serves as an effective tracer of stellar ages, previous studies have employed a variety of models, mostly first-, second-, or third-degree polynomials, to describe this relation. In this work, we systematically examine the chromospheric activity--age relation across different stellar types by using a large combined sample of stars from open clusters and the field. The LAMOST spectroscopic data are used to calculate the chromospheric activity indices.

We model the activity--age relation using first-, second-, and third-degree polynomials, compute the Bayesian \emph{evidence} for each model, and apply Jeffreys’ scale to select the most appropriate model.
Our analysis indicates that a linear (Skumanich-type) relation between log$_{10}(R_{\rm{HK}}^{'})$ and log$_{10}({\rm Age})$ provides a best description of how stellar chromospheric activity declines with age, with the slope varying across different spectral types. 
However, we also observe variations in the decay rate of chromospheric activity with age, which require more accurate follow-up investigation.
In addition, although the Sun itself is among the least active stars, many solar twins exhibit activity levels similar to that of the present-day Sun.
Finally, our results show that lower-metallicity stars exhibit enhanced activity for F-, G-, and K-type stars, consistent with weaker \cahk absorption. However, no clear metallicity dependence is observed for M dwarfs, since their strong TiO molecular bands will overwhelm the already quite faint \cahk absorption.

\begin{acknowledgements}
We thank the referee for the comprehensive comments and suggestions, which have greatly improved this manuscript. The Guoshoujing Telescope (the Large Sky Area Multi-Object Fiber Spectroscopic Telescope LAMOST) is a National Major Scientific Project built by the Chinese Academy of Sciences. Funding for the project has been provided by the National Development and Reform Commission. LAMOST is operated and managed by the National Astronomical Observatories, Chinese Academy of Sciences. 
This work was supported by National Natural Science Foundation of China (NSFC) under grant Nos. 12588202/12273057/11833002/12090042, the National Key Research and Development Program of China (NKRDPC) under grant number 2023YFA1607901, the Strategic Priority Program of the Chinese Academy of Sciences under grant number XDB1160302, and science research grants from the China Manned Space Project. J.F.L acknowledges the support from the New Cornerstone Science Foundation through the New Cornerstone Investigator Program and the XPLORER PRIZE.
\end{acknowledgements}

\bibliography{main}{}
\bibliographystyle{aasjournalv7}

\clearpage
\begin{appendix}
\section{Fitting results}
\label{sec:app}
\renewcommand\thefigure{\Alph{section}\arabic{figure}}
\renewcommand\thetable{\Alph{section}\arabic{table}}
\setcounter{figure}{0}
\setcounter{table}{0}  

\begin{figure*}[h]
\centering
\subfigure[]{
\includegraphics[width=0.45\textwidth]{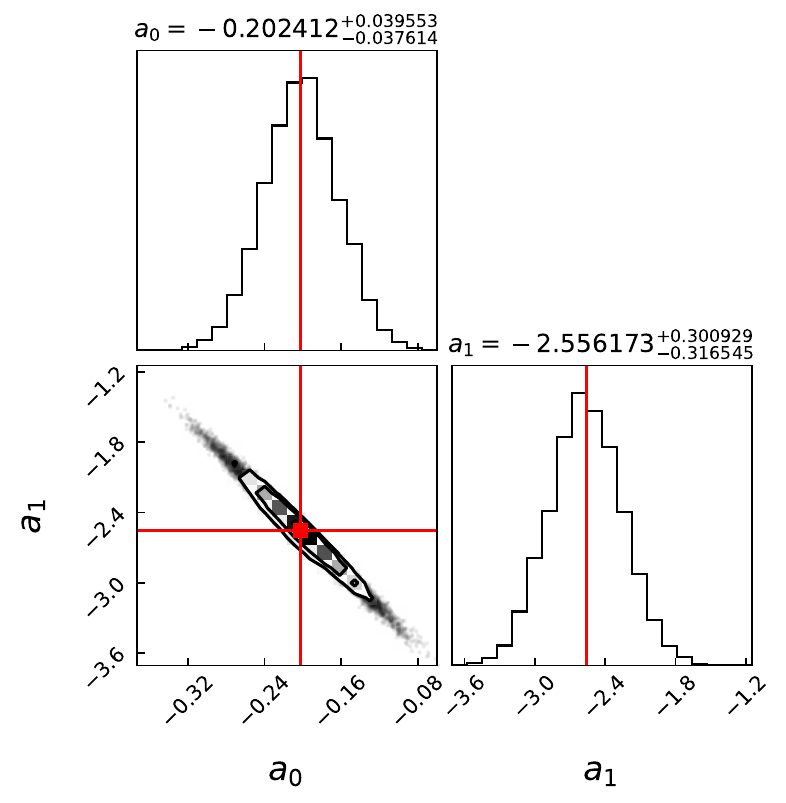}}
\subfigure[]{
\includegraphics[width=0.45\textwidth]{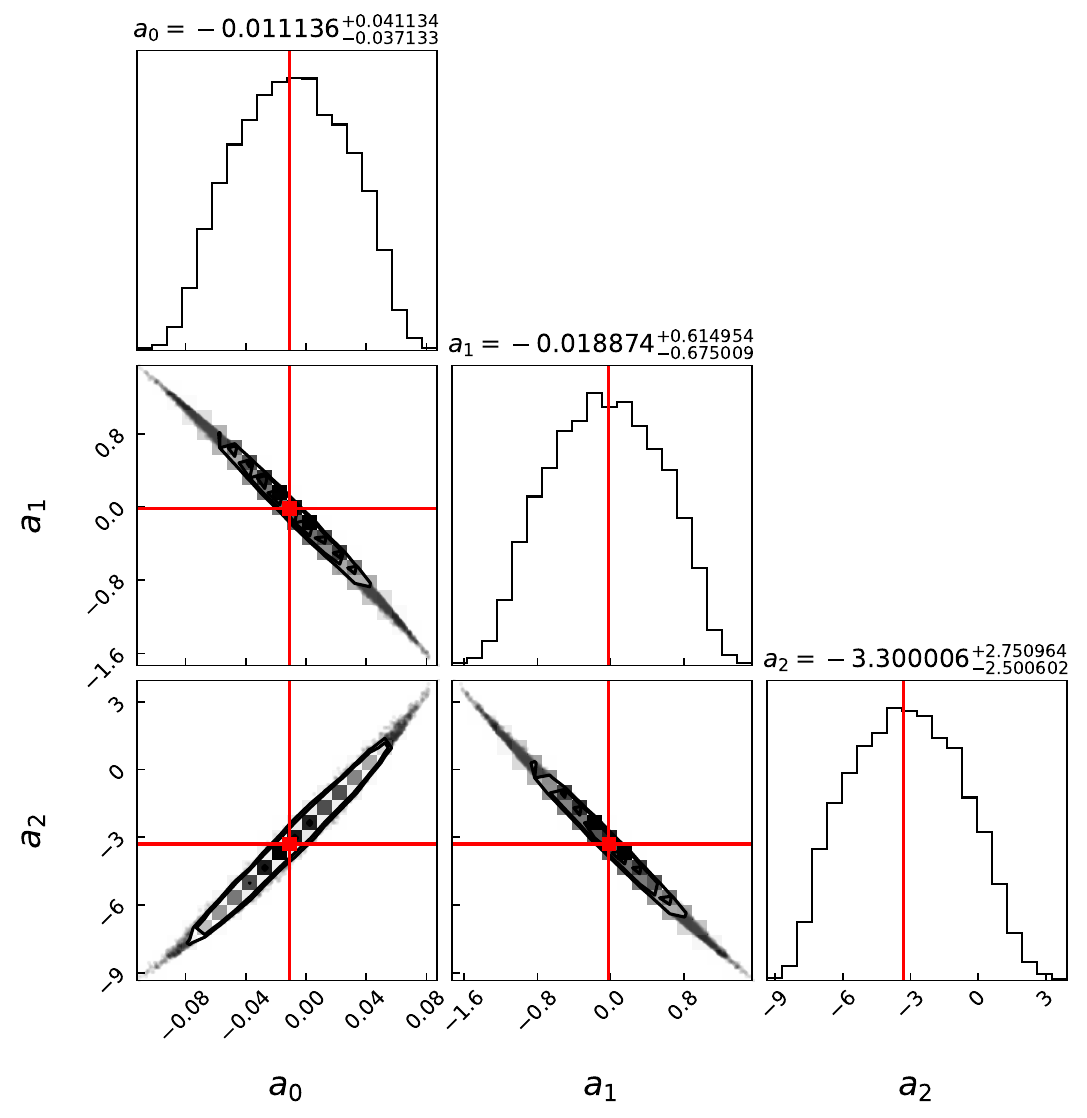}}
\subfigure[]{
\includegraphics[width=0.45\textwidth]{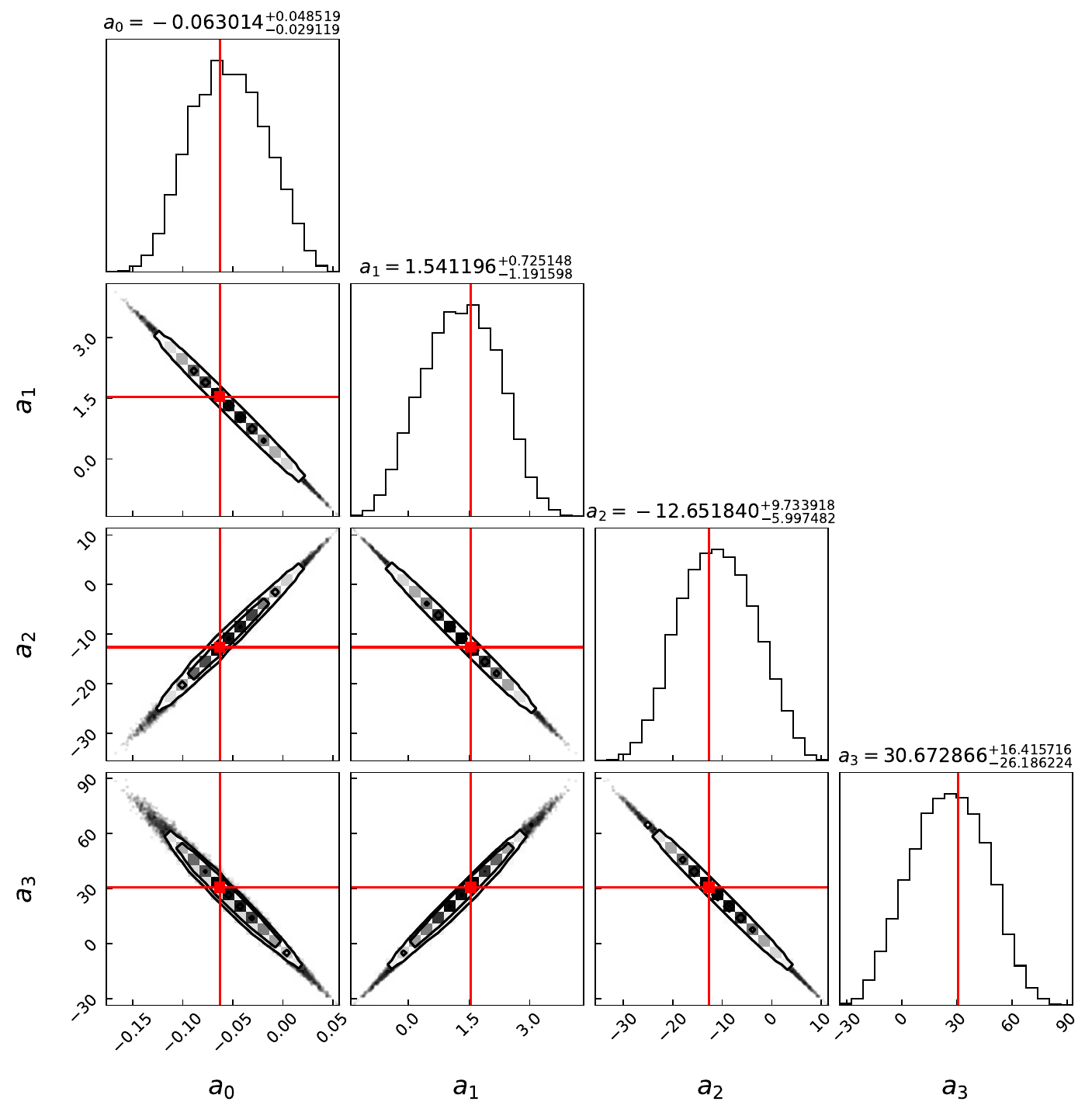}}

\caption{Examples of posterior probability distributions for the G‑type star fits, using the standard deviation of log$_{10}(R_{\rm{HK}}^{'})$ in each age bin as uncertainty. Panel (a), (b), and (c) represents 1st-, 2nd-, and 3rd-degree polynomial model, respectively.}
\label{corner.fig}
\end{figure*}

\clearpage

\begin{figure*}[h]
\centering
\subfigure[]{
\includegraphics[width=0.45\textwidth]{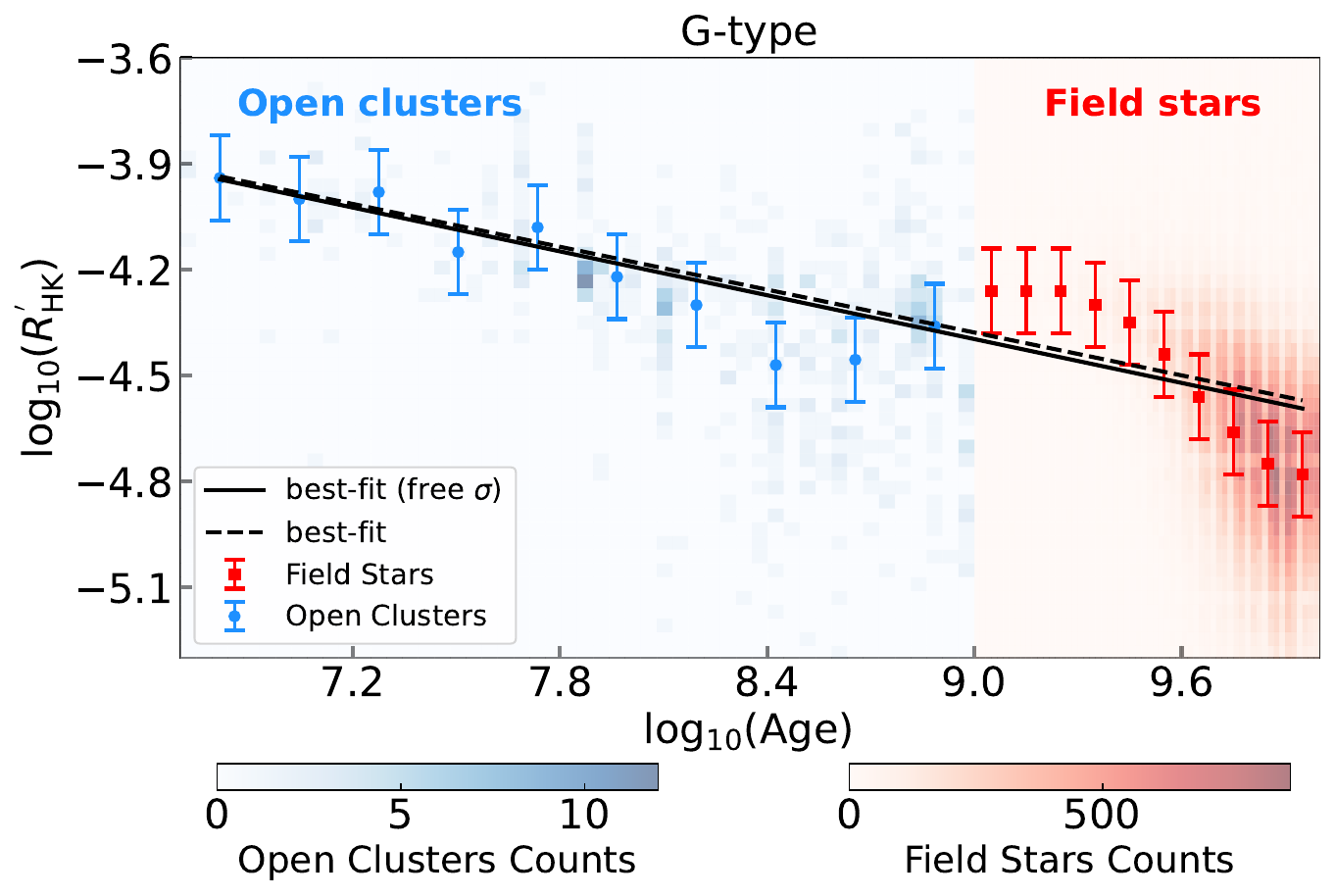}}
\subfigure[]{
\includegraphics[width=0.45\textwidth]{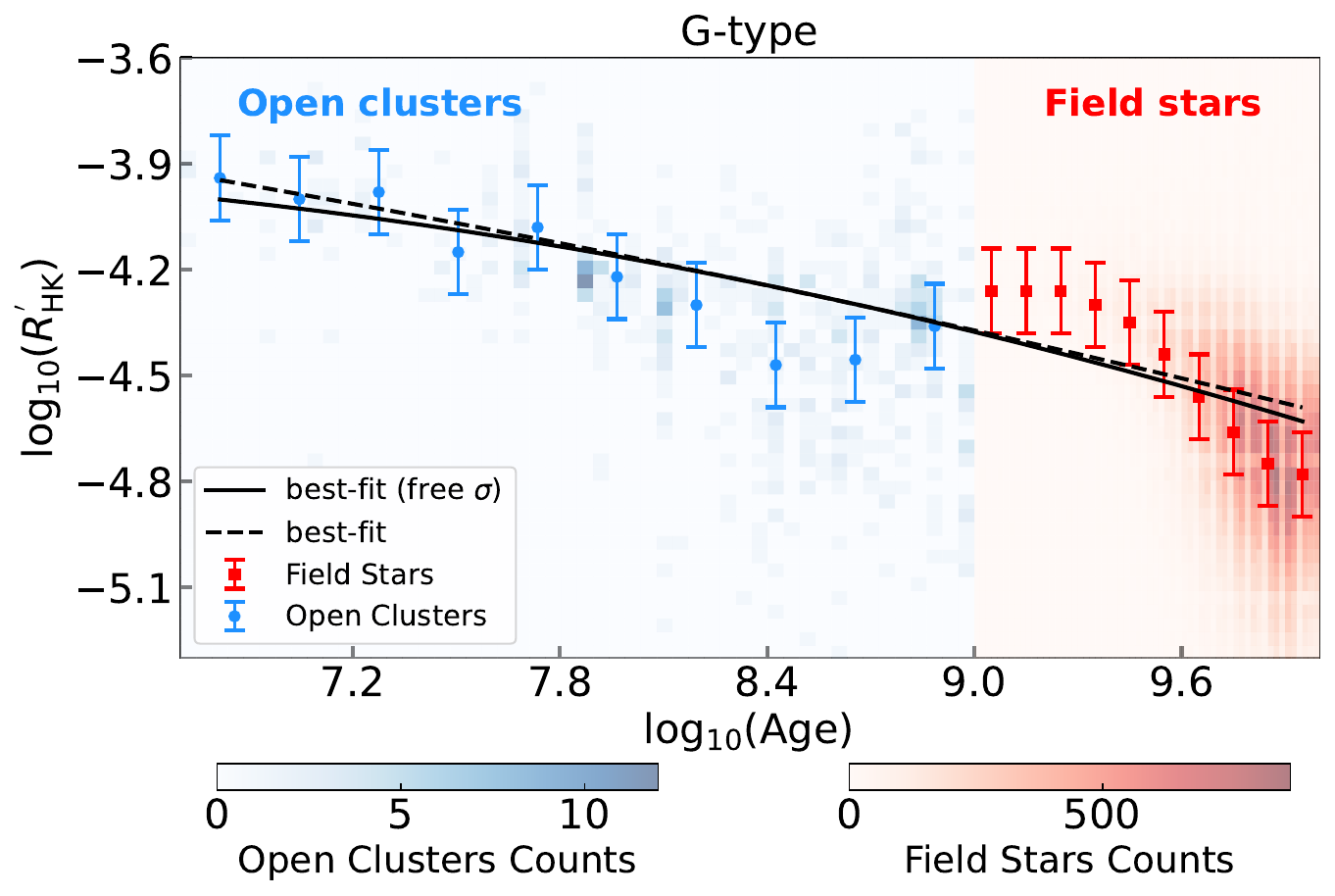}}
\centering
\subfigure[]{
\includegraphics[width=0.45\textwidth]{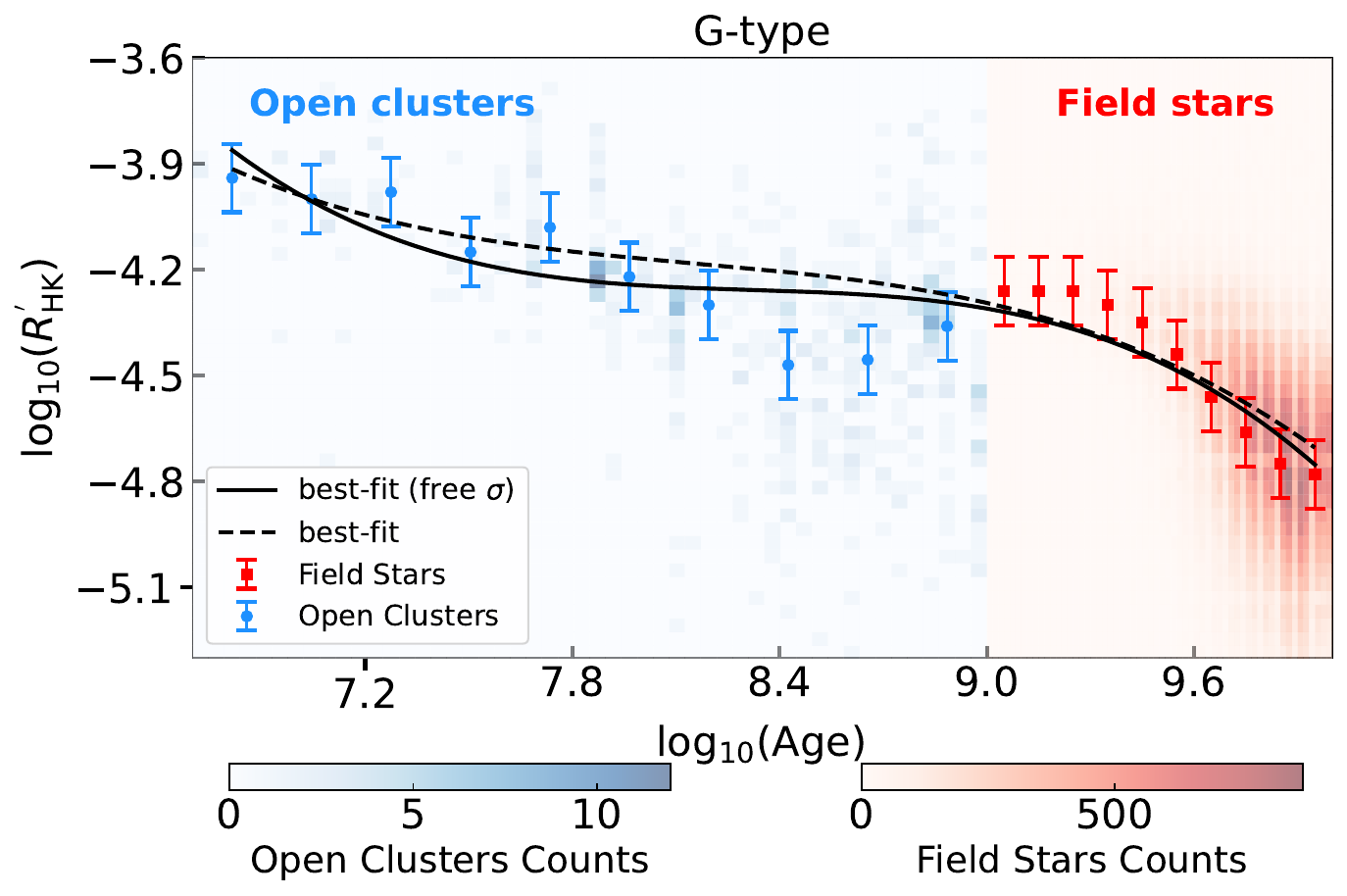}}
\caption{Comparison of $R_{\rm{HK}}^{'}$--age relations for G-type stars with different methods: fits using the standard deviation of log$_{10}(R_{\rm{HK}}^{'})$ in each bin as the error (dashed lines), and fits with the error of log$_{10}(R_{\rm{HK}}^{'})$ set as a free parameter (solid lines).
Panels (a), (b), and (c) show the results for first‑, second‑, and third‑order polynomial models, respectively.}
\label{comparison.fig}
\end{figure*}

\begin{figure*}[h]
\centering
\subfigure[]{
\includegraphics[width=0.45\textwidth]{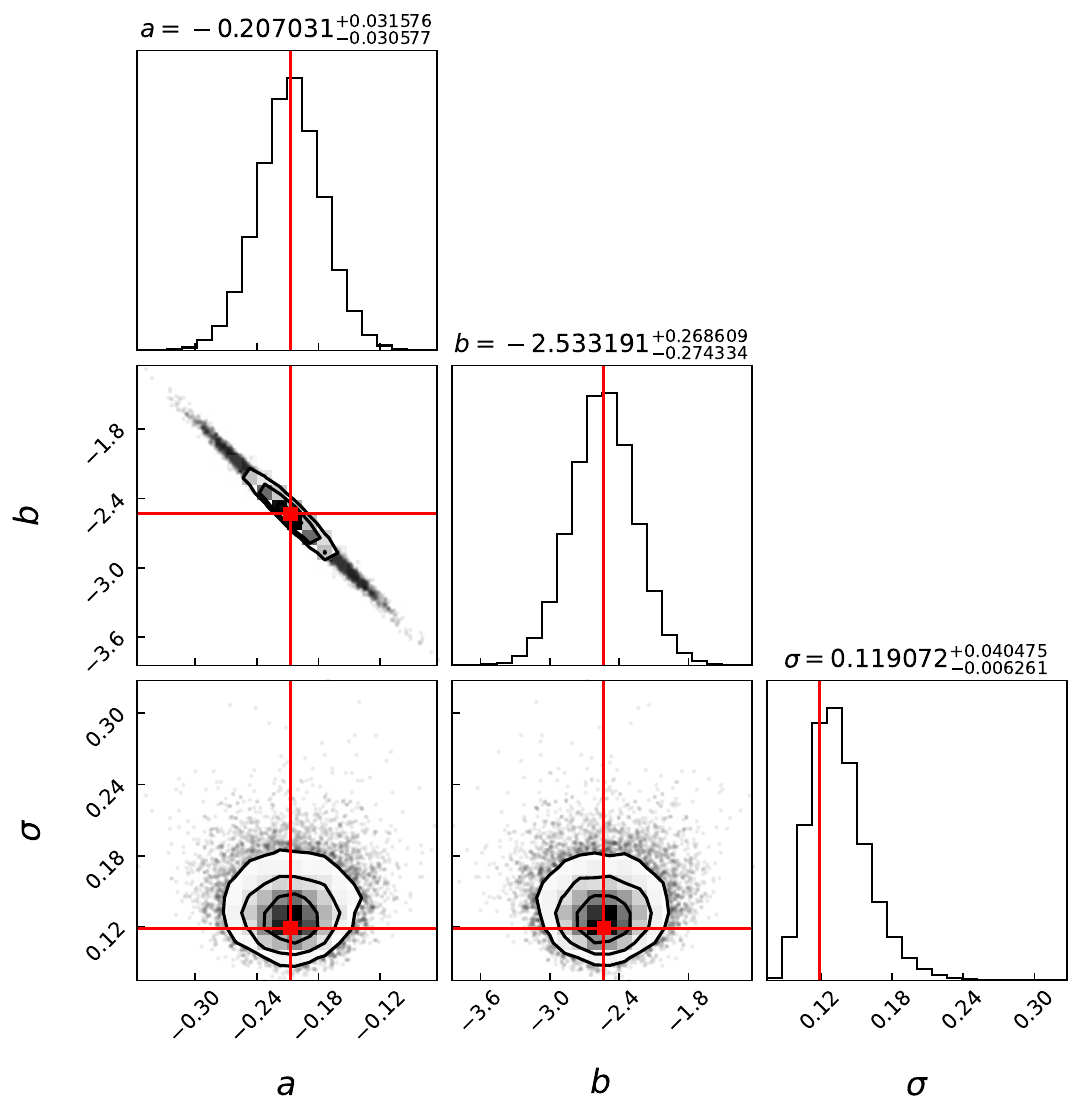}}
\subfigure[]{
\includegraphics[width=0.45\textwidth]{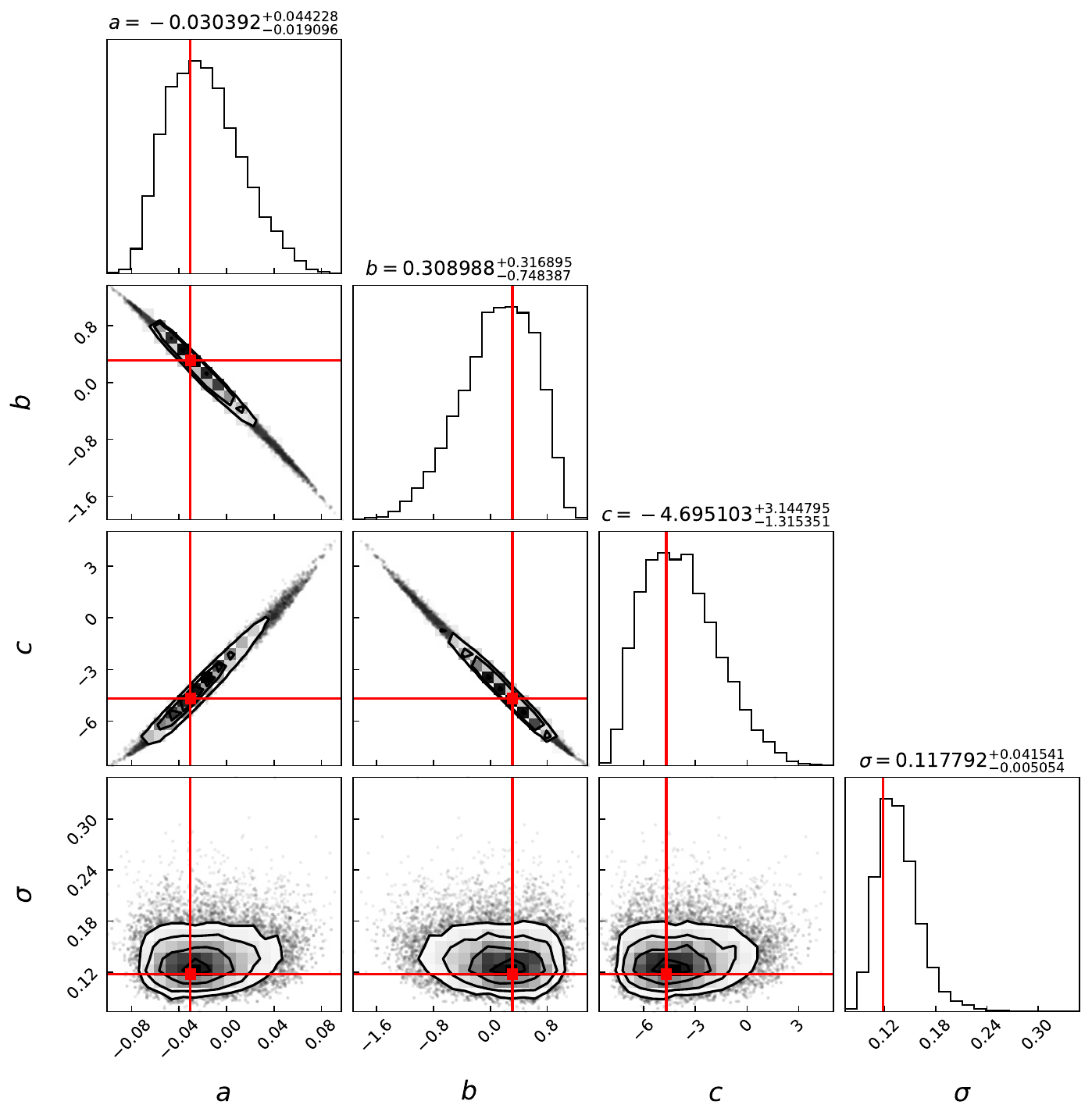}}
\centering
\subfigure[]{
\includegraphics[width=0.45\textwidth]{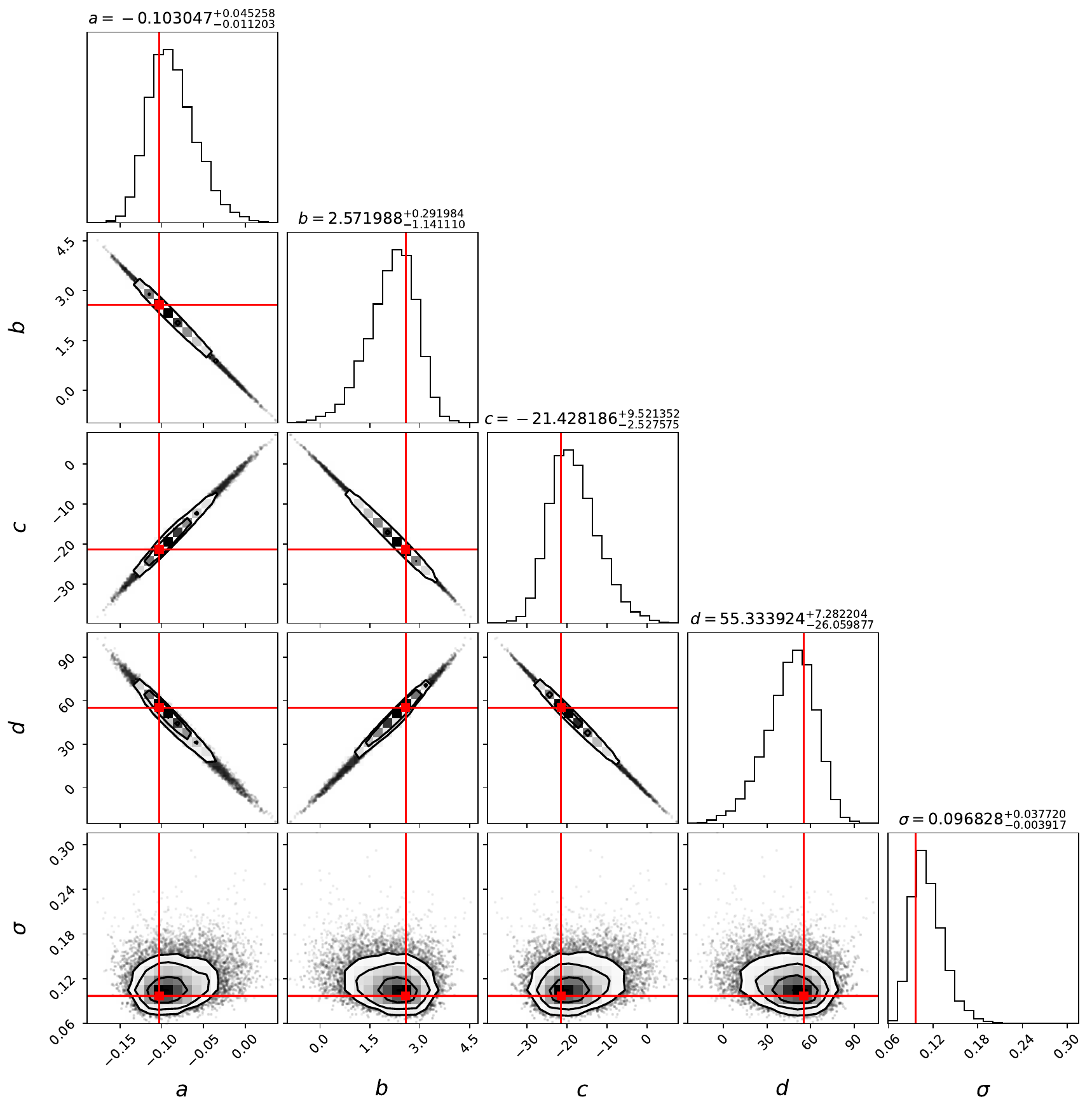}}
\caption{Posterior probability distributions for the G‑type star fits, using $\sigma$ as a free parameter. Panel (a), (b), and (c) represents 1st-, 2nd-, and 3rd-degree polynomial model, respectively.}
\label{comparison2.fig}
\end{figure*}

\clearpage

\end{appendix}



\end{document}